\documentclass{JHEP3}
\usepackage{amsmath}\allowdisplaybreaks[1]
\usepackage{graphicx}

\newcommand{\kslash}{\rlap{$\hspace{.06ex}/$}k}
\newcommand{\Dslash}{\rlap{$\hspace{.38ex}/$}\Delta}

\title{Generalized parton correlation functions for a spin-1/2 hadron}
\author{Stephan Mei{\ss}ner\\
 Institut f{\"u}r Theoretische Physik II, Ruhr-Universit{\"a}t Bochum,\\
 44780 Bochum, Germany\\
 E-mail: \email{stephan.meissner@tp2.rub.de}}
\author{Andreas Metz\\
 Department of Physics, Temple University,\\
 Philadelphia, PA 19122-6082, U.S.A.\\
 E-mail: \email{metza@temple.edu}}
\author{Marc Schlegel\\
 Theory Center, Jefferson Lab, 12000 Jefferson Avenue,\\
 Newport News, VA 23606, U.S.A.\\
 E-mail: \email{schlegel@jlab.org}}

\abstract{The fully unintegrated, off-diagonal quark-quark correlator for a 
spin-1/2 hadron is parameterized in terms of so-called generalized parton 
correlation functions.
Such objects, in particular, can be considered as {\it mother distributions} 
of generalized parton distributions on the one hand and transverse momentum 
dependent parton distributions on the other.
Therefore, our study provides new, model-independent insights into the 
recently proposed nontrivial relations between generalized and transverse 
momentum dependent parton distributions.
We find that none of these relations can be promoted to a model-independent 
status.
As a by-product we obtain the first complete classification of generalized 
parton distributions beyond leading twist.
The present paper is a natural extension of our previous corresponding analysis 
for spin-0 hadrons.}
\preprint{}
\begin{document}

%
%
%
\section{Introduction}\label{c:sec1}
In a recent work we parameterized the fully unintegrated, off-diagonal quark-quark 
correlator for a spin-0 hadron in terms of so-called generalized parton 
correlation functions (GPCFs)~\cite{Meissner:2008ay}.
The GPCFs depend on the full 4-momentum of the quark and, in addition, on the 
momentum transfer to the hadron.
As such they contain the maximum amount of information about the partonic 
structure of hadrons.
The purpose of the present paper is to extend this analysis to the more
interesting but at the same time more challenging case of a spin-1/2 hadron.
Related work on the (simpler) unintegrated diagonal quark-quark correlator for 
a spin-1/2 hadron can be found in 
refs.~\cite{Goeke:2005hb,Collins:2007ph,Rogers:2008jk}.

GPCFs are of particular interest because of their connection to the generalized 
parton distributions 
(GPDs)~\cite{Mueller:1998fv,Ji:1996ek,Radyushkin:1996nd,Goeke:2001tz,Diehl:2003ny,
Belitsky:2005qn,Boffi:2007yc} and the transverse momentum dependent parton
distributions (TMDs) \cite{Mulders:1995dh, Barone:2001sp, Bacchetta:2006tn,D'Alesio:2007jt}. 
Both GPDs and TMDs have been intensely studied during the last 15 years.
While GPDs appear in the QCD-description of hard exclusive reactions such as
deep virtual Compton scattering or hard exclusive meson production, TMDs can be
measured in certain semi-inclusive reactions like semi-inclusive deep inelastic
scattering (SIDIS) or the Drell-Yan (DY) process. 
These two types of parton distributions provide a 3-dimensional picture 
of the nucleon --- either in a mixed position-momentum representation or in
pure momentum space.
Moreover, they contain important information on the orbital motion of
partons inside the nucleon.
The important point is that both the GPDs and the TMDs appear as two 
different limiting cases of the GPCFs.
Therefore, the GPCFs can be considered as {\it mother distributions} of 
GPDs and TMDs~\cite{Ji:2003ak,Belitsky:2003nz,Belitsky:2005qn}.
Note that the GPCFs also have a direct connection to the so-called Wigner 
distributions --- the quantum mechanical analogues of classical phase space 
distributions --- of the hadron-parton 
system~\cite{Ji:2003ak,Belitsky:2003nz,Belitsky:2005qn}. 

In the present paper, as the major application of the classification of the 
GPCFs, we obtain new, model-independent information on the nontrivial 
relations between GPDs and TMDs which have been suggested in the 
literature~\cite{Burkardt:2002ks,Burkardt:2003uw,Burkardt:2003je,Diehl:2005jf,
Burkardt:2005hp,Lu:2006kt,Meissner:2007rx,Pasquini:2008ax}. 
In order to study this point we exploit the connection between the GPCFs on 
the one hand as well as the GPDs and TMDs on the other, and explore, 
in particular, which GPDs and TMDs have the same {\it mother distributions}.
The nontrivial relations between GPDs and TMDs attracted a lot of attention
during the last years.
The most prominent case, first proposed in ref.~\cite{Burkardt:2002ks}, 
is the relation between the so-called Sivers 
TMD~\cite{Sivers:1989cc,Sivers:1990fh} and the GPD $E$. 
This connection provides a rather intuitive understanding of the Sivers 
single spin asymmetry in SIDIS which has been explored by the HERMES and the 
COMPASS 
experiments~\cite{Airapetian:2004tw,Alexakhin:2005iw,Ageev:2006da,Collaboration:2009ti}.
Although in the meantime various nontrivial relations between GPDs and TMDs 
were established in simple models (see~\cite{Meissner:2007rx} for an 
overview and~\cite{Pasquini:2008ax}), no model-independent relations have 
been obtained so far.
In fact, our previous work on GPCFs showed that for spin-0 hadrons no 
model-independent relations between GPDs and TMDs can be established.
In the present work we arrive at the same conclusion for spin-1/2 
hadrons.
A first account on the spin-1/2 case can be found in the conference 
contribution~\cite{Meissner:2008xs}.

If the GPCFs are integrated upon one light-cone component of the quark 
momentum one arrives at the so-called generalized transverse momentum 
dependent parton distributions (GTMDs) which can show up in the 
description of hard exclusive reactions.
While quark GTMDs typically appear at subleading twist --- and in cases
where the standard collinear factorization cannot be applied ---
(see, e.g., refs.~\cite{Vanderhaeghen:1999xj,Diehl:2007hd,Goloskokov:2007nt}), 
gluon GTMDs have been extensively used to describe processes at high 
energies (low $x$) like, for instance, diffractive vector 
meson~\cite{Martin:1999wb} and Higgs production at the Tevatron and
the LHC~\cite{Khoze:2000cy,Albrow:2008pn,Martin:2009ku} in the framework 
of the so-called $k_T$ factorization.
Also an approximate method for (theoretically) constraining the 
unpolarized gluon GTMD has been proposed~\cite{Martin:2001ms}.
In the present work we will not further elaborate on the phenomenology 
of GTMDs, although it is an important topic (for related work see also 
refs.~\cite{Collins:2007ph,Rogers:2008jk}).

The plan of the manuscript is as follows.
In the next section the parameterization of the generalized quark-quark 
correlator for a spin-1/2 hadron in terms of GPCFs is presented.
This parameterization forms the basis for the rest of the paper.
In section~\ref{c:sec3} we consider the GTMDs.
The results in that section follow in a straightforward way from those
in section~\ref{c:sec2}.
The TMD-limit and the GPD-limit for the GTMDs are investigated in 
section~\ref{c:sec4}, providing us with the first complete counting of 
GPDs beyond leading twist.
In particular, we also explore which GPDs and TMDs have the same 
{\it mother distributions}.
The outcome of this analysis allows us to investigate the model-independent 
status of possible nontrivial relations between GPDs and TMDs.
Section \ref{c:sec5} contains the conclusions.
Details of the (technically demanding) derivation of the classification for
the GPCFs can be found in appendix~\ref{c:app_par}.
The exact relations between the GPCFs and the GTMDs defined in the manuscript 
are given in appendix~\ref{c:app_gtmd_gpcf}, while in 
appendix~\ref{c:app_gtmd_model} our model-independent study is supplemented by 
the calculation of the leading twist GTMDs in a simple diquark spectator model 
for the nucleon.

%
%
%
\section{Generalized parton correlation functions}\label{c:sec2}

%
%
%
\subsection{Definition}
In this section we derive the structure of the generalized, fully-unintegrated 
quark-quark correlator for a spin-1/2 hadron which is defined as
\begin{equation}
 W_{\lambda \lambda'}^{[\Gamma]}(P, k, \Delta, N; \eta)
 = \frac{1}{2} \int \frac{d^4 z}{(2\pi)^4} \, e^{i k \cdot z} \, \langle p', \lambda' | \,
   \bar{\psi}(-\tfrac{1}{2}z) \, \Gamma \, \mathcal{W}(-\tfrac{1}{2}z, \tfrac{1}{2}z \, | \, n) \,
   \psi(\tfrac{1}{2}z) \, | p, \lambda \rangle \,.
 \label{e:corr_gpcf}
\end{equation}
The correlator $W$ depends on the helicities $\lambda$ and $\lambda'$, the average 
momentum $P = (p+p')/2$ of the initial and final hadron, the momentum transfer 
$\Delta = p' - p$ to the hadron, and the average quark momentum $k$. 
(For the kinematics we also refer to figure~\ref{f:kinematics}.) 
The object $\Gamma$ is an element of the complete basis 
$\{1, \gamma_5, \gamma^\mu, \gamma^\mu\gamma_5, i\sigma^{\mu\nu}\}$ 
with $\sigma^{\mu\nu} = i [\gamma^{\mu},\gamma^{\nu}] / 2$.
The Wilson line $\mathcal{W}$ ensures the color gauge invariance of the correlator 
in eq.~(\ref{e:corr_gpcf}) and is running along the path\footnote{The path of the
Wilson line is chosen such that appropriate Wilson lines are obtained when taking
the GPD-limit and the TMD-limit (see also section 2.4).}
\begin{equation}
 -\tfrac{1}{2}z \;\to\; -\tfrac{1}{2}z + \infty \cdot n 
 \;\to\; \tfrac{1}{2}z + \infty \cdot n \;\to\; \tfrac{1}{2}z \,, 
 \label{e:path}
\end{equation}
with all four points connected by straight lines. 
It is now important to realize that the integration contour of the Wilson line not 
only depends on the coordinates of the initial and final points but also on the 
light-cone direction which is opposite to the direction of $P$~\cite{Goeke:2003az}. 
This induces a dependence on a light-cone vector $n$.  
In fact, instead of using $n$ a rescaled vector $\lambda n$ with some positive 
parameter $\lambda$ could be taken in order to specify the Wilson line. 
Therefore, the correlator actually only depends on the vector
\begin{equation}
 N = \frac{M^2 \, n}{P \cdot n} \,,
 \label{e:direction}
\end{equation}
which is invariant under the mentioned rescaling. 
For convenience in~(\ref{e:direction}) the hadron mass $M$ is used such that $N$ 
has the same mass dimension as an ordinary 4-momentum. 
The parameter $\eta$ in~(\ref{e:corr_gpcf}) is defined through the zeroth component 
of $n$ according to
\begin{equation}
 \eta = \text{sign}(n_0) \,,
\end{equation} 
which means that we simultaneously treat future-pointing $(\eta = +1)$ and 
past-pointing $(\eta = - 1)$ Wilson lines. 
Keeping this dependence is particularly convenient once we make the projection of 
the correlator in~(\ref{e:corr_gpcf}) onto the correlator defining TMDs.
%
\FIGURE[t]{%
 \includegraphics{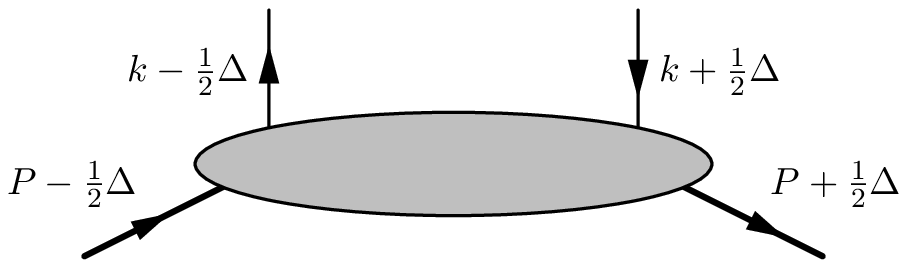}
 \caption{Kinematics for GPCFs.}
 \label{f:kinematics}}
%

%
%
%
\subsection{Parameterization}
In order to obtain the parameterization of the correlator in~(\ref{e:corr_gpcf}) in terms 
of GPCFs it is necessary to analyze its behavior under parity.
One finds that
\begin{align}
 &W_{\lambda \lambda'}^{[\Gamma]}(P, k, \Delta, N; \eta) \nonumber\\
 &\quad = \frac{1}{2} \int \frac{d^4 z}{(2\pi)^4} \,
  e^{i k \cdot z} \, \langle p', \lambda' | \, \hat{P}^\dagger \hat{P} \,
  \bar{\psi}(-\tfrac{1}{2}z) \, \hat{P}^\dagger \hat{P} \, \Gamma \, \hat{P}^\dagger \hat{P} \,
  \mathcal{W}(-\tfrac{1}{2}z, \tfrac{1}{2}z \, | \, n) \, \hat{P}^\dagger \hat{P} \, \psi(\tfrac{1}{2}z) \,
  \hat{P}^\dagger \hat{P} \, | p, \lambda \rangle \nonumber\\
 &\quad = \frac{1}{2} \int \frac{d^4 z}{(2\pi)^4} \, e^{i k \cdot z} \, \langle \bar{p}', \lambda_P' | \,
  \bar{\psi}(-\tfrac{1}{2}\bar{z}) \, \gamma_0 \, \Gamma \, \gamma_0 \,
  \mathcal{W}(-\tfrac{1}{2}\bar{z}, \tfrac{1}{2}\bar{z} \, | \, \bar{n}) \,
  \psi(\tfrac{1}{2}\bar{z}) \, | \bar{p}, \lambda_P \rangle \nonumber\\
 &\quad = \frac{1}{2} \int \frac{d^4 z}{(2\pi)^4} \, e^{i \bar{k} \cdot z} \, \langle \bar{p}', \lambda_P' | \,
  \bar{\psi}(-\tfrac{1}{2}z) \, \gamma_0 \, \Gamma \, \gamma_0 \,
  \mathcal{W}(-\tfrac{1}{2}z, \tfrac{1}{2}z \, | \, \bar{n}) \,
  \psi(\tfrac{1}{2}z) \, | \bar{p}, \lambda_P \rangle \nonumber\\
 &\quad = W_{\lambda_P \lambda_P'}^{[\gamma_0 \, \Gamma \, \gamma_0]}
  (\bar{P}, \bar{k}, \bar{\Delta}, \bar{N}; \eta) \,,
  \label{e:par}
\end{align}
where $\bar{P}^\mu = P_\mu = (P^0,-\vec{P})$ etc., while $\lambda_P$ and $\lambda_P'$ 
denote the parity-reversed helicities $\lambda$ and $\lambda'$.
We now introduce the (dimensionless) matrix functions $\Gamma_\text{S}$, 
$\Gamma_\text{P}$, $\Gamma^\mu_\text{V}$, $\Gamma^\mu_\text{A}$, 
and $\Gamma^{\mu\nu}_\text{T}$ through
\begin{align}
 W_{\lambda \lambda'}^{[1]}(P, k, \Delta, N; \eta)
 &= \bar{u}(p', \lambda') \, \Gamma_\text{S}(P, k, \Delta, N; \eta) \, u(p, \lambda)
 && \text{(scalar)}\label{e:sdb} \,,\\
 W_{\lambda \lambda'}^{[\gamma_5]}(P, k, \Delta, N; \eta)
 &= \bar{u}(p', \lambda') \, \Gamma_\text{P}(P, k, \Delta, N; \eta) \, u(p, \lambda)
 && \text{(pseudoscalar)} \label{e:pdb} \,,\\
 W_{\lambda \lambda'}^{[\gamma^\mu]}(P, k, \Delta, N; \eta)
 &= \bar{u}(p', \lambda') \, \Gamma^\mu_\text{V}(P, k, \Delta, N; \eta) \, u(p, \lambda)
 && \text{(vector)} \label{e:vdb} \,,\\
 W_{\lambda \lambda'}^{[\gamma^\mu \gamma_5]}(P, k, \Delta, N; \eta)
 &= \bar{u}(p', \lambda') \, \Gamma^\mu_\text{A}(P, k, \Delta, N; \eta) \, u(p, \lambda)
 && \text{(axial vector)} \label{e:adb} \,,\\
 W_{\lambda \lambda'}^{[i\sigma^{\mu\nu}]}(P, k, \Delta, N; \eta)
 &= \bar{u}(p', \lambda') \, \Gamma^{\mu\nu}_\text{T}(P, k, \Delta, N; \eta) \, u(p, \lambda)
 && \text{(tensor)} \label{e:tdb} \,.
\end{align}
From eq.~(\ref{e:par}) it follows for the scalar matrix function in eq.~(\ref{e:sdb})
\begin{align}
 &\bar{u}(p', \lambda') \, \Gamma_\text{S}(P, k, \Delta, N; \eta) \, u(p, \lambda) \nonumber\\
 &\ =\bar{u}(\bar{p}', \lambda_P') \,
  \Gamma_\text{S}(\bar{P}, \bar{k}, \bar{\Delta}, \bar{N}; \eta) \, u(\bar{p}, \lambda_P) \nonumber\\
 &\ =\bar{u}(p', \lambda') \, \hat{P}^\dagger \,
  \Gamma_\text{S}(\bar{P}, \bar{k}, \bar{\Delta}, \bar{N}; \eta) \,
  \hat{P} \, u(p, \lambda) \nonumber\\
 &\ =\bar{u}(p', \lambda') \, \gamma_0 \,
  \Gamma_\text{S}(\bar{P}, \bar{k}, \bar{\Delta}, \bar{N}; \eta) \,
  \gamma_0 \, u(p, \lambda) \,.
  \label{e:par2}
\end{align}
Analogous results hold for the other matrix functions in eqs.~(\ref{e:pdb})--(\ref{e:tdb}),
and one finds
\begin{eqnarray}
 \Gamma_\text{S}(P, k, \Delta, N; \eta) &=&
 + \gamma_0 \, \Gamma_\text{S}(\bar{P}, \bar{k}, \bar{\Delta}, \bar{N}; \eta) \, \gamma_0 \,,
 \label{e:ws_parity}\\
 \Gamma_\text{P}(P, k, \Delta, N; \eta) &=&
 - \gamma_0 \, \Gamma_\text{P}(\bar{P}, \bar{k}, \bar{\Delta}, \bar{N}; \eta) \, \gamma_0 \,,
 \label{e:wp_parity}\\
 \Gamma^\mu_\text{V}(P, k, \Delta, N; \eta) &=&
 + \gamma_0 \, \Gamma^{\bar{\mu}}_\text{V}(\bar{P}, \bar{k}, \bar{\Delta}, \bar{N}; \eta) \, \gamma_0 \,,
 \label{e:wv_parity}\\
 \Gamma^\mu_\text{A}(P, k, \Delta, N; \eta) &=&
 - \gamma_0 \, \Gamma^{\bar{\mu}}_\text{A}(\bar{P}, \bar{k}, \bar{\Delta}, \bar{N}; \eta) \, \gamma_0 \,,
 \label{e:wa_parity}\\
 \Gamma^{\mu\nu}_\text{T}(P, k, \Delta, N; \eta) &=&
 + \gamma_0 \, \Gamma^{\bar{\mu}\bar{\nu}}_\text{T}(\bar{P}, \bar{k}, \bar{\Delta}, \bar{N}; \eta) \, \gamma_0
 \label{e:wt_parity}
\end{eqnarray}
for their behavior under parity.
It turns out that the general structure of the correlator $W$ can already be 
obtained on the basis of the parity constraints 
in~(\ref{e:ws_parity})--(\ref{e:wt_parity}).
One ends up with 64 linearly independent matrix structures multiplied by scalar 
functions (for the derivation see appendix~\ref{c:app_par}),
\begin{align}
 &W_{\lambda \lambda'}^{[1]}(P, k, \Delta, N; \eta) \nonumber\\*
 &\quad = \bar{u}(p', \lambda') \, \bigg[
  A^E_{1}
  + \frac{i\sigma^{k\Delta}}{M^2} \, A^E_{2}
  + \frac{i\sigma^{kN}}{M^2} \, A^E_{3}
  + \frac{i\sigma^{\Delta N}}{M^2} \, A^E_{4} \bigg] \, u(p, \lambda) \,,
 \label{e:ws_res}\\
 &W_{\lambda \lambda'}^{[\gamma_5]}(P, k, \Delta, N; \eta) \nonumber\\*
 &\quad = \bar{u}(p', \lambda') \, \bigg[
  \frac{i\varepsilon^{Pk\Delta N}}{M^4} \, A^E_{5}
  + \frac{i\sigma^{PN} \gamma_5}{M^2} \, A^E_{6}
  + \frac{i\sigma^{kN} \gamma_5}{M^2} \, A^E_{7}
  + \frac{i\sigma^{\Delta N} \gamma_5}{M^2} \, A^E_{8} \bigg] \, u(p, \lambda) \,,
 \label{e:wp_res}\\
 &W_{\lambda \lambda'}^{[\gamma^\mu]}(P, k, \Delta, N; \eta) \nonumber\\*
 &\quad = \bar{u}(p', \lambda') \, \bigg[
  \frac{P^\mu}{M} \, A^F_{1}
  + \frac{k^\mu}{M} \, A^F_{2}
  + \frac{\Delta^\mu}{M} \, A^F_{3}
  + \frac{N^\mu}{M} \, A^F_{4}
  + \frac{i\sigma^{\mu k}}{M} \, A^F_{5}
  + \frac{i\sigma^{\mu \Delta}}{M} \, A^F_{6}
  + \frac{i\sigma^{\mu N}}{M} \, A^F_{7}
  \nonumber\\*
 &\quad\hspace{2.55ex}
  + \frac{P^\mu \, i\sigma^{k\Delta}}{M^3} \, A^F_{8}
  + \frac{k^\mu \, i\sigma^{k\Delta}}{M^3} \, A^F_{9}
  + \frac{N^\mu \, i\sigma^{k\Delta}}{M^3} \, A^F_{10}
  + \frac{P^\mu \, i\sigma^{kN}}{M^3} \, A^F_{11}
  + \frac{k^\mu \, i\sigma^{kN}}{M^3} \, A^F_{12}
  \nonumber\\*
 &\quad\hspace{2.55ex}
  + \frac{N^\mu \, i\sigma^{kN}}{M^3} \, A^F_{13}
  + \frac{P^\mu \, i\sigma^{\Delta N}}{M^3} \, A^F_{14}
  + \frac{\Delta^\mu \, i\sigma^{\Delta N}}{M^3} \, A^F_{15}
  + \frac{N^\mu \, i\sigma^{\Delta N}}{M^3} \, A^F_{16}
 \bigg] \, u(p, \lambda) \,,
 \label{e:wv_res}\\
 &W_{\lambda \lambda'}^{[\gamma^\mu \gamma_5]}(P, k, \Delta, N; \eta) \nonumber\\*
 &\quad = \bar{u}(p', \lambda') \, \bigg[
  \frac{i\varepsilon^{\mu Pk\Delta}}{M^3} \, A^G_{1}
  + \frac{i\varepsilon^{\mu PkN}}{M^3} \, A^G_{2}
  + \frac{i\varepsilon^{\mu P\Delta N}}{M^3} \, A^G_{3}
  + \frac{i\varepsilon^{\mu k\Delta N}}{M^3} \, A^G_{4}
  \nonumber\\*
 &\quad\hspace{2.55ex}
  + \frac{i\sigma^{\mu P} \gamma_5}{M} \, A^G_{5}
  + \frac{i\sigma^{\mu k} \gamma_5}{M} \, A^G_{6}
  + \frac{i\sigma^{\mu N} \gamma_5}{M} \, A^G_{7}
  + \frac{k^\mu \, i\sigma^{PN} \gamma_5}{M^3} \, A^G_{8}
  + \frac{\Delta^\mu \, i\sigma^{PN} \gamma_5}{M^3} \, A^G_{9}
  \nonumber\\*
 &\quad\hspace{2.55ex}
  + \frac{N^\mu \, i\sigma^{PN} \gamma_5}{M^3} \, A^G_{10}
  + \frac{k^\mu \, i\sigma^{kN} \gamma_5}{M^3} \, A^G_{11}
  + \frac{\Delta^\mu \, i\sigma^{kN} \gamma_5}{M^3} \, A^G_{12}
  + \frac{N^\mu \, i\sigma^{kN} \gamma_5}{M^3} \, A^G_{13}
  \nonumber\\*
 &\quad\hspace{2.55ex}
  + \frac{P^\mu \, i\sigma^{\Delta N} \gamma_5}{M^3} \, A^G_{14}
  + \frac{\Delta^\mu \, i\sigma^{\Delta N} \gamma_5}{M^3} \, A^G_{15}
  + \frac{N^\mu \, i\sigma^{\Delta N} \gamma_5}{M^3} \, A^G_{16}
 \bigg] \, u(p, \lambda) \,,
 \label{e:wa_res}\\
 &W_{\lambda \lambda'}^{[i\sigma^{\mu\nu}]}(P, k, \Delta, N; \eta) \nonumber\\*
 &\quad = (\delta^\mu_\rho \delta^\nu_\sigma - \delta^\nu_\rho \delta^\mu_\sigma) \, \bar{u}(p', \lambda') \, \bigg[
  \frac{P^\rho k^\sigma}{M^2} \, A^H_{1}
  + \frac{P^\rho \Delta^\sigma}{M^2} \, A^H_{2}
  + \frac{P^\rho N^\sigma}{M^2} \, A^H_{3}
  + \frac{k^\rho \Delta^\sigma}{M^2} \, A^H_{4}
  \nonumber\\*
 &\quad\hspace{2.55ex}
  + \frac{k^\rho N^\sigma}{M^2} \, A^H_{5}
  + \frac{\Delta^\rho N^\sigma}{M^2} \, A^H_{6}
  + i\sigma^{\rho\sigma} \, A^H_{7}
  + \frac{P^\rho \, i\sigma^{\sigma k}}{M^2} \, A^H_{8}
  + \frac{N^\rho \, i\sigma^{\sigma k}}{M^2} \, A^H_{9}
  \nonumber\\*
 &\quad\hspace{2.55ex}
  + \frac{P^\rho \, i\sigma^{\sigma\Delta}}{M^2} \, A^H_{10}
  + \frac{N^\rho \, i\sigma^{\sigma\Delta}}{M^2} \, A^H_{11}
  + \frac{P^\rho \, i\sigma^{\sigma N}}{M^2} \, A^H_{12}
  + \frac{k^\rho \, i\sigma^{\sigma N}}{M^2} \, A^H_{13}
  + \frac{\Delta^\rho \, i\sigma^{\sigma N}}{M^2} \, A^H_{14}
  \nonumber\\*
 &\quad\hspace{2.55ex}
  + \frac{N^\rho \, i\sigma^{\sigma N}}{M^2} \, A^H_{15}
  + \frac{P^\rho k^\sigma \, i\sigma^{k\Delta}}{M^4} \, A^H_{16}
  + \frac{P^\rho N^\sigma \, i\sigma^{k\Delta}}{M^4} \, A^H_{17}
  + \frac{k^\rho N^\sigma \, i\sigma^{k\Delta}}{M^4} \, A^H_{18}
  \nonumber\\*
 &\quad\hspace{2.55ex}
  + \frac{P^\rho k^\sigma \, i\sigma^{kN}}{M^4} \, A^H_{19}
  + \frac{P^\rho N^\sigma \, i\sigma^{kN}}{M^4} \, A^H_{20}
  + \frac{k^\rho N^\sigma \, i\sigma^{kN}}{M^4} \, A^H_{21}
  + \frac{P^\rho \Delta^\sigma \, i\sigma^{\Delta N}}{M^4} \, A^H_{22}
  \nonumber\\*
 &\quad\hspace{2.55ex}
  + \frac{P^\rho N^\sigma \, i\sigma^{\Delta N}}{M^4} \, A^H_{23}
  + \frac{\Delta^\rho N^\sigma \, i\sigma^{\Delta N}}{M^4} \, A^H_{24}
 \bigg] \, u(p, \lambda) \,,
 \label{e:wt_res}
\end{align}
where we used 
$\varepsilon^{abcd} = \varepsilon^{\mu\nu\rho\sigma} a_\mu b_\nu c_\rho d_\sigma$ and 
$\sigma^{ab} = \sigma^{\mu\nu} a_\mu b_\nu$ to shorten the notation.
Our treatment leading to~(\ref{e:ws_res})--(\ref{e:wt_res}) is analogous to what has 
already been done for a spin-0 hadron~\cite{Meissner:2008ay}.
The functions $A^E_i$, $A^F_i$, $A^G_i$, and $A^H_i$ are independent and represent 
the GPCFs.
They depend on all possible scalar products of the momenta $P$, $k$, $\Delta$, and $N$ 
as well as the parameter $\eta$.
The various factors of $M$ are introduced in order to assign the same mass dimension 
to all GPCFs.
Note that the parameterizations~(\ref{e:ws_res})--(\ref{e:wt_res}) are ambiguous in 
the sense that one can always rewrite them into other forms by means of the Gordon 
identities~(\ref{e:gi1})--(\ref{e:gi4}).
However, the amount of structures as presented in 
eqs.~(\ref{e:ws_res})--(\ref{e:wt_res}) is minimized.
For further details we refer to appendix~\ref{c:app_par}.

%
%
%
\subsection{Properties}
By applying hermiticity and time reversal to the correlator in~(\ref{e:corr_gpcf}) it 
is possible to derive some basic properties of the GPCFs.
From hermiticity it follows that
\begin{align}
 &\Big[ W_{\lambda \lambda'}^{[\Gamma]}(P, k, \Delta, N; \eta) \Big]^* \nonumber\\
 &\quad = \frac{1}{2} \int \frac{d^4 z}{(2\pi)^4} \,
  e^{-i k \cdot z} \, \langle p', \lambda' | \, \bar{\psi}(-\tfrac{1}{2}z) \,
  \Gamma \, \mathcal{W}(-\tfrac{1}{2}z, \tfrac{1}{2}z \, | \, n) \, \psi(\tfrac{1}{2}z) \, | p, \lambda \rangle^*
  \nonumber\\
 &\quad = \frac{1}{2} \int \frac{d^4 z}{(2\pi)^4} \,
  e^{-i k \cdot z} \, \langle p, \lambda | \, \bar{\psi}(\tfrac{1}{2}z) \,
  \gamma_0 \, \Gamma^\dagger \, \gamma_0 \,
  \mathcal{W}(\tfrac{1}{2}z, -\tfrac{1}{2}z \, | \, n) \, \psi(-\tfrac{1}{2}z) \, | p', \lambda' \rangle
  \nonumber\\
 &\quad = \frac{1}{2} \int \frac{d^4 z}{(2\pi)^4} \,
  e^{i k \cdot z} \, \langle p, \lambda | \, \bar{\psi}(-\tfrac{1}{2}z) \,
  \gamma_0 \, \Gamma^\dagger \, \gamma_0 \,
  \mathcal{W}(-\tfrac{1}{2}z, \tfrac{1}{2}z \, | \, n) \, \psi(\tfrac{1}{2}z) \, | p', \lambda' \rangle
  \nonumber\\
 &\quad = W_{\lambda' \lambda}^{[\gamma_0 \, \Gamma^\dagger \, \gamma_0]}(P, k, -\Delta, N; \eta) \,.
\end{align}
For the matrix functions in eqs.~(\ref{e:sdb})--(\ref{e:tdb}) this leads to
\begin{eqnarray}
 \Big[ \Gamma_\text{S}(P, k, \Delta, N; \eta) \Big]^\dagger
 &=& + \gamma_0 \, \Gamma_\text{S}(P, k, -\Delta, N; \eta) \, \gamma_0 \,,
 \label{e:ws_hermiticity}\\
 \Big[ \Gamma_\text{P}(P, k, \Delta, N; \eta) \Big]^\dagger
 &=& - \gamma_0 \, \Gamma_\text{P}(P, k, -\Delta, N; \eta) \, \gamma_0 \,,
 \label{e:wp_hermiticity}\\
 \Big[ \Gamma^\mu_\text{V}(P, k, \Delta, N; \eta) \Big]^\dagger
 &=& + \gamma_0 \, \Gamma^\mu_\text{V}(P, k, -\Delta, N; \eta) \, \gamma_0 \,,
 \label{e:wv_hermiticity}\\
 \Big[ \Gamma^\mu_\text{A}(P, k, \Delta, N; \eta) \Big]^\dagger
 &=& + \gamma_0 \, \Gamma^\mu_\text{A}(P, k, -\Delta, N; \eta) \, \gamma_0 \,,
 \label{e:wa_hermiticity}\\
 \Big[ \Gamma^{\mu\nu}_\text{T}(P, k, \Delta, N; \eta) \Big]^\dagger
 &=& - \gamma_0 \, \Gamma^{\mu\nu}_\text{T}(P, k, -\Delta, N; \eta) \, \gamma_0 \,.
 \label{e:wt_hermiticity}
\end{eqnarray}
Applying the hermiticity constraints~(\ref{e:ws_hermiticity})--(\ref{e:wt_hermiticity}) 
to the decomposition in~(\ref{e:ws_res})--(\ref{e:wt_res}) one finds
\begin{equation}
 X^*(P, k, \Delta, N; \eta) = \pm X(P, k, -\Delta, N; \eta) \,,
 \label{e:gpcf_hermiticity}
\end{equation}
where the plus sign holds for $X = A^E_{1}$, $A^E_{2}$, $A^E_{4}$, $A^E_{8}$, $A^F_{1}$, 
$A^F_{2}$, $A^F_{4}$, $A^F_{6}$, $A^F_{8}$, $A^F_{9}$, $A^F_{10}$, $A^F_{14}$, $A^F_{16}$, 
$A^G_{1}$, $A^G_{3}$, $A^G_{4}$, $A^G_{5}$, $A^G_{6}$, $A^G_{7}$, $A^G_{8}$, $A^G_{10}$, 
$A^G_{11}$, $A^G_{13}$, $A^G_{15}$, $A^H_{2}$, $A^H_{4}$, $A^H_{6}$, $A^H_{7}$, $A^H_{8}$, 
$A^H_{9}$, $A^H_{12}$, $A^H_{13}$, $A^H_{15}$, $A^H_{19}$, $A^H_{20}$, $A^H_{21}$, 
$A^H_{22}$, $A^H_{24}$ and the minus sign for all the other GPCFs. 

In addition, time reversal leads to
\begin{align}
 &\Big[ W_{\lambda \lambda'}^{[\Gamma]}(P, k, \Delta, N; \eta) \Big]^* \nonumber\\
 &\quad = \frac{1}{2} \int \frac{d^4 z}{(2\pi)^4} \,
  e^{-i k \cdot z} \, \langle p', \lambda' | \, \bar{\psi}(-\tfrac{1}{2}z) \,
  \Gamma \, \mathcal{W}(-\tfrac{1}{2}z, \tfrac{1}{2}z \, | \, n) \, \psi(\tfrac{1}{2}z) \, | p, \lambda \rangle^*
  \nonumber\\
 &\quad = \frac{1}{2} \int \frac{d^4 z}{(2\pi)^4} \,
  e^{-i k \cdot z} \, \langle p', \lambda' | \, \hat{T}^\dagger \hat{T} \,
  \bar{\psi}(-\tfrac{1}{2}z) \, \hat{T}^\dagger \hat{T} \, \Gamma \, \hat{T}^\dagger \hat{T} \,
  \mathcal{W}(-\tfrac{1}{2}z, \tfrac{1}{2}z \, | \, n) \, \hat{T}^\dagger \hat{T} \, \psi(\tfrac{1}{2}z) \,
  \hat{T}^\dagger \hat{T} \, | p, \lambda \rangle \nonumber\\
 &\quad = \frac{1}{2} \int \frac{d^4 z}{(2\pi)^4} \, e^{-i k \cdot z} \, \langle \bar{p}', \lambda_T' | \,
  \bar{\psi}(\tfrac{1}{2}\bar{z}) \, (-i\gamma_5 C) \, \Gamma^* \, (-i\gamma_5 C) \,
  \mathcal{W}(\tfrac{1}{2}\bar{z}, -\tfrac{1}{2}\bar{z} \, | \, -\bar{n}) \,
  \psi(-\tfrac{1}{2}\bar{z}) \, | \bar{p}, \lambda_T \rangle \nonumber\\
 &\quad = \frac{1}{2} \int \frac{d^4 z}{(2\pi)^4} \, e^{i \bar{k} \cdot z} \, \langle \bar{p}', \lambda_T' | \,
  \bar{\psi}(-\tfrac{1}{2}z) \, (-i\gamma_5 C) \, \Gamma^* \, (-i\gamma_5 C) \,
  \mathcal{W}(-\tfrac{1}{2}z, \tfrac{1}{2}z \, | \, -\bar{n}) \,
  \psi(\tfrac{1}{2}z) \, | \bar{p}, \lambda_T \rangle \nonumber\\
 &\quad = W_{\lambda_T \lambda_T'}^{[(-i\gamma_5 C) \, \Gamma^* \, (-i\gamma_5 C)]}
  (\bar{P}, \bar{k}, \bar{\Delta}, \bar{N}; -\eta) \,,
\end{align}
where $C$ is the charge conjugation matrix, while $\lambda_T$ and $\lambda_T'$ denote 
the time-reversed helicities $\lambda$ and $\lambda'$. 
Analogous to eq.~(\ref{e:par2}) one finds for the matrix functions in 
eqs.~(\ref{e:sdb})--(\ref{e:tdb})
\begin{eqnarray}
 \Big[ \Gamma_\text{S}(P, k, \Delta, N; \eta) \Big]^*
 &=& (-i\gamma_5 C) \, \Gamma_\text{S}(\bar{P}, \bar{k}, \bar{\Delta}, \bar{N}; -\eta) \,
     (-i\gamma_5 C) \,,
 \label{e:ws_timereversal}\\
 \Big[ \Gamma_\text{P}(P, k, \Delta, N; \eta) \Big]^*
 &=& (-i\gamma_5 C) \, \Gamma_\text{P}(\bar{P}, \bar{k}, \bar{\Delta}, \bar{N}; -\eta) \,
     (-i\gamma_5 C) \,,
 \label{e:wp_timereversal}\\
 \Big[ \Gamma^\mu_\text{V}(P, k, \Delta, N; \eta) \Big]^*
 &=& (-i\gamma_5 C) \, \Gamma^{\bar{\mu}}_\text{V}(\bar{P}, \bar{k}, \bar{\Delta}, \bar{N}; -\eta) \,
     (-i\gamma_5 C) \,,
 \label{e:wv_timereversal}\\
 \Big[ \Gamma^\mu_\text{A}(P, k, \Delta, N; \eta) \Big]^*
 &=& (-i\gamma_5 C) \, \Gamma^{\bar{\mu}}_\text{A}(\bar{P}, \bar{k}, \bar{\Delta}, \bar{N}; -\eta) \,
     (-i\gamma_5 C) \,,
 \label{e:wa_timereversal}\\
 \Big[ \Gamma^{\mu\nu}_\text{T}(P, k, \Delta, N; \eta) \Big]^*
 &=& (-i\gamma_5 C) \, \Gamma^{\bar{\mu}\bar{\nu}}_\text{T}(\bar{P}, \bar{k}, \bar{\Delta}, \bar{N}; -\eta) \,
     (-i\gamma_5 C) \,.
 \label{e:wt_timereversal}
\end{eqnarray}
The time-reversal constraints~(\ref{e:ws_timereversal})--(\ref{e:wt_timereversal}) provide  
\begin{equation}
 X^*(P, k, \Delta, N; \eta) = X(P, k, \Delta, N; -\eta)
 \label{e:gpcf_timereversal}
\end{equation}
for all GPCFs, relating those defined with future-pointing Wilson lines to those 
defined with past-pointing lines.

From these considerations it follows that in general GPCFs, unlike GPDs or TMDs, are 
complex-valued functions. 
Keeping now in mind that $\eta \in \{-1\, ,1\}$ and using eq.~(\ref{e:gpcf_timereversal}) 
one finds immediately that only the imaginary part of the GPCFs depends on $\eta$. 
This allows one to write
\begin{equation}
 X(P, k, \Delta, N; \eta) = X^{e}(P, k, \Delta, N) + i \, X^{o}(P, k, \Delta, N; \eta) \,,  
 \label{e:gpcf_decomp}
\end{equation}
with
\begin{equation}
 X^{o}(P, k, \Delta, N; \eta) = - X^{o}(P, k, \Delta, N; -\eta) \,,
 \label{e:gpcf_sign}
\end{equation}
where we call $X^{e}$ the T-even and $X^{o}$ the T-odd part of the generic GPCF $X$. 
The sign reversal of $X^{o}$ in eq.~(\ref{e:gpcf_sign}) when going from 
future-pointing to past-pointing Wilson lines corresponds to the sign reversal 
discussed in ref.~\cite{Collins:2002kn} for T-odd TMDs.

%
%
%
\subsection{Limits}
Now we would like to give a first account on the relation between GPCFs on the one 
hand and GPDs as well as TMDs on the other. 
To this end we consider the quark-quark correlator $F$ defining GPDs for a spin-1/2 
target, which can be obtained from the correlator $W$ in eq.~(\ref{e:corr_gpcf}) 
by means of the projection
\begin{align}
 & F_{\lambda \lambda'}^{[\Gamma]}(P, x, \Delta, N)
   = \int dk^- \, d^2\vec{k}_T \, W_{\lambda \lambda'}^{[\Gamma]}(P, k, \Delta, N; \eta) \nonumber \\
 & \quad = \frac{1}{2}\int \frac{dz^-}{2 \pi} \, e^{i k \cdot z} \,
   \langle p', \lambda' | \, \bar{\psi}(-\tfrac{1}{2}z) \, \Gamma \,
   {\cal W}(-\tfrac{1}{2}z,\tfrac{1}{2}z\,|\,n) \,
   \psi(\tfrac{1}{2}z) \, | p, \lambda \rangle \, \Big|_{z^+ = \vec{z}_T = 0} \,.
 \label{e:corr_gpd}
\end{align}
In this formula we use light-cone components that are specified through
$a^{\pm}=(a^0\pm a^3)/\sqrt{2}$ and $\vec{a}_T = (a^1,a^2)$ for a generic 
4-vector $a = (a^0,a^1,a^2,a^3)$, where, in particular, we choose $k^+ = x P^+$. 
Note that after integrating upon $k^-$ and $\vec{k}_T$ the dependence on the parameter 
$\eta$ drops out. 
It is well-known that in this case we are dealing with a light-cone correlator and 
the two quark fields are just connected by a straight line. 
This means that the choice of the contour in~(\ref{e:path}) leads, after projection, 
to the appropriate Wilson line for the GPD-correlator.

The correlator $\Phi$ defining TMDs can be extracted from $W$ by putting 
$\Delta = 0$ and integrating out one light-cone component of the quark momentum 
(which we choose to be $k^-$),
\begin{align}
 & \Phi_{\lambda \lambda'}^{[\Gamma]}(P, x, \vec{k}_T, N; \eta)
   = \int dk^- \, W_{\lambda \lambda'}^{[\Gamma]}(P, k, 0, N; \eta) \nonumber \\
 & \quad = \frac{1}{2}\int \frac{dz^- \, d^2 \vec{z}_T}{(2\pi)^3} \, e^{i k \cdot z} \,
   \langle P, \lambda' | \, \bar{\psi}(-\tfrac{1}{2}z) \, \Gamma \,
   {\cal W}(-\tfrac{1}{2}z,\tfrac{1}{2}z\,|\,n) \,
   \psi(\tfrac{1}{2}z) \, | P, \lambda \rangle \, \Big|_{z^+ = 0} \,.
 \label{e:corr_tmd}
\end{align}
Note that for $\Delta = 0$ one has $p = p^{\prime} = P$. 
We point out that the path specified in~(\ref{e:path}) also leads to a proper 
Wilson line after taking the TMD-limit~\cite{Collins:1981uw,Collins:1999dz,Collins:2000gd,Collins:2002kn,Ji:2002aa,Belitsky:2002sm,Collins:2004nx,Cherednikov:2007tw,Cherednikov:2008ua}. 
Since $\Phi$ in eq.~(\ref{e:corr_tmd}) is not a light-cone correlator the dependence 
on the parameter $\eta$ remains. 
The case $\eta = +1$ is appropriate for defining TMDs in processes with final state 
interactions of the struck quark like SIDIS, while $\eta = -1$ can be used for TMDs 
in DY~\cite{Collins:2002kn}. 
It has been emphasized in 
refs.~\cite{Collins:1981uw,Collins:2003fm,Hautmann:2007uw,Collins:2008ht} that, 
in general, light-like Wilson lines as used in the unintegrated correlators in 
(\ref{e:corr_gpcf}) and (\ref{e:corr_tmd}) lead to divergences. 
Such divergences can be avoided, however, by adopting a near light-cone direction. 
For the purpose of the present work it is sufficient to note that our general reasoning 
remains valid once a near light-cone direction is used instead of $n$.

It is evident that not only the correlators $F$ and $\Phi$ appear as projections of 
the most general two-parton correlator $W$ as outlined above, but also the GPDs and 
the TMDs are projections of certain GPCFs. 
Therefore, GPCFs can be considered as {\it mother distributions}, which actually 
contain the maximum amount of information on the two-parton structure of 
hadrons~\cite{Ji:2003ak,Belitsky:2003nz,Belitsky:2005qn}. 
Despite this fact a classification of the GPCFs as given 
in~(\ref{e:ws_res})--(\ref{e:wt_res}) has never been worked out.

%
%
%
\section{Generalized transverse momentum dependent parton distributions}\label{c:sec3}

%
%
%
\subsection{Definition}
The projections in~(\ref{e:corr_gpd}) and~(\ref{e:corr_tmd}) contain the integration upon 
the minus-component of the quark momentum. 
Therefore, it is useful to consider in more detail the correlator
\begin{align}
 & W_{\lambda \lambda'}^{[\Gamma]}(P, x, \vec{k}_T, \Delta, N; \eta)
   = \int dk^- \, W_{\lambda \lambda'}^{[\Gamma]}(P, k, \Delta, N; \eta) \nonumber \\
 & \quad = \frac{1}{2}\int \frac{dz^- \, d^2 \vec{z}_T}{(2\pi)^3} \, e^{i k \cdot z} \,
   \langle p', \lambda' | \, \bar{\psi}(-\tfrac{1}{2}z) \, \Gamma \,
   {\cal W}(-\tfrac{1}{2}z,\tfrac{1}{2}z\,|\,n) \,
   \psi(\tfrac{1}{2}z) \, | p, \lambda \rangle \, \Big|_{z^+ = 0} \,.
 \label{e:corr_gtmd}
\end{align}
Below the parameterization of this object is given in terms of what we call generalized 
transverse momentum dependent parton distributions (GTMDs). 
Of course, this result can now be obtained in a straightforward manner on the basis of 
the decomposition in eqs.~(\ref{e:ws_res})--(\ref{e:wt_res}). 
On the basis of the above discussion it is obvious that also the GTMDs, like the GPCFs, 
can be considered as {\it mother distributions} of GPDs and TMDs. 
It is the correlator in~(\ref{e:corr_gtmd}) which for instance can enter the description 
of hard exclusive meson production~\cite{Goloskokov:2007nt}, while the corresponding 
correlator for gluons appears when considering diffractive processes in lepton-hadron 
as well as hadron-hadron 
collisions~\cite{Martin:1999wb,Khoze:2000cy,Albrow:2008pn,Martin:2009ku}. 
The question whether or not it appears with a Wilson line as defined in~(\ref{e:path}) 
to our knowledge has never been addressed in the literature and requires further 
investigation that goes beyond the scope of the present work.

For our analysis we choose an infinite momentum frame such that $P$ has a large 
plus-momentum and no transverse momentum. 
The plus-component of $\Delta$ is expressed through the commonly used variable $\xi$. 
To be now precise the 4-momenta in~(\ref{e:ws_res})--(\ref{e:wt_res}) are specified 
according to
\begin{eqnarray}
 P & = & \bigg[\, P^+ \, , \, \frac{\vec{\Delta}_T^2 + 4M^2}{8(1-\xi^2)P^+} \, , \, \vec{0}_T \, \bigg] \,, \\
 k & = & \bigg[\, x P^+ \, ,\, k^- \, , \, \vec{k}_T \, \bigg] \,, \\
 \Delta & = & \bigg[\, -2 \xi P^+ \, , \, \frac{\xi\vec{\Delta}_T^2 + 4\xi M^2}{4(1-\xi^2)P^+} \, , \, 
                    \vec{\Delta}_T \, \bigg] \,, \\
 n & = & \bigg[\, 0\, , \, \pm 1 \, , \, \vec{0}_T \, \bigg] \,.
 \label{e:lcvec}
\end{eqnarray} 
The vector $n$ in eq.~(\ref{e:lcvec}) is of course not the most general light-cone 
vector. 
In particular, it has no transverse component and points opposite to the direction 
of $P$ as already mentioned earlier. 
However, if one wants to arrive at an appropriate definition of TMDs for SIDIS and DY, 
there is no freedom left for this vector because it is fixed by the external momenta 
of the respective processes.

%
%
%
\subsection{Parameterization}
Now we have all the ingredients which are needed for writing down the final result for 
the generalized $k_T$-dependent correlator~(\ref{e:corr_gtmd}) in terms of GTMDs. 
We start with the twist-2 case for which one gets
\begin{eqnarray}
 W_{\lambda \lambda'}^{[\gamma^+]}
 &=& \frac{1}{2M} \, \bar{u}(p', \lambda') \, \bigg[
      F_{1,1}
      + \frac{i\sigma^{i+} k_T^i}{P^+} \, F_{1,2}
      + \frac{i\sigma^{i+} \Delta_T^i}{P^+} \, F_{1,3} \nonumber\\*
 & &  + \frac{i\sigma^{ij} k_T^i \Delta_T^j}{M^2} \, F_{1,4}
     \bigg] \, u(p, \lambda)
     \,, \label{e:gtmd_1}\\
 W_{\lambda \lambda'}^{[\gamma^+\gamma_5]}
 &=& \frac{1}{2M} \, \bar{u}(p', \lambda') \, \bigg[
      - \frac{i\varepsilon_T^{ij} k_T^i \Delta_T^j}{M^2} \, G_{1,1}
      + \frac{i\sigma^{i+}\gamma_5 k_T^i}{P^+} \, G_{1,2}
      + \frac{i\sigma^{i+}\gamma_5 \Delta_T^i}{P^+} \, G_{1,3} \nonumber\\*
 & &  + i\sigma^{+-}\gamma_5 \, G_{1,4}
     \bigg] \, u(p, \lambda)
     \,, \label{e:gtmd_2}\\
 W_{\lambda \lambda'}^{[i\sigma^{j+}\gamma_5]}
 &=& \frac{1}{2M} \, \bar{u}(p', \lambda') \, \bigg[
      - \frac{i\varepsilon_T^{ij} k_T^i}{M} \, H_{1,1}
      - \frac{i\varepsilon_T^{ij} \Delta_T^i}{M} \, H_{1,2}
      + \frac{M \, i\sigma^{j+}\gamma_5}{P^+} \, H_{1,3} \nonumber\\*
 & &  + \frac{k_T^j \, i\sigma^{k+}\gamma_5 k_T^k}{M \, P^+} \, H_{1,4}
      + \frac{\Delta_T^j \, i\sigma^{k+}\gamma_5 k_T^k}{M \, P^+} \, H_{1,5}
      + \frac{\Delta_T^j \, i\sigma^{k+}\gamma_5 \Delta_T^k}{M \, P^+} \, H_{1,6} \nonumber\\*
 & &  + \frac{k_T^j \, i\sigma^{+-}\gamma_5}{M} \, H_{1,7}
      + \frac{\Delta_T^j \, i\sigma^{+-}\gamma_5}{M} \, H_{1,8}
     \bigg] \, u(p, \lambda)
     \,. \label{e:gtmd_3}
\end{eqnarray}
Here the definitions $\varepsilon^{0123} = 1$ and 
$\varepsilon_T^{ij} = \varepsilon^{-+ij}$ are used. 
The 16 complex-valued twist-2 GTMDs $F_{1,i}$, $G_{1,i}$, and $H_{1,i}$ are given by 
$k^{-}$-integrals of certain linear combinations of the GPCFs 
in~(\ref{e:wv_res})--(\ref{e:wt_res}), where the explicit relations are listed in 
appendix~\ref{c:app_gtmd_gpcf}. 
To shorten the notation the arguments on both sides of the 
eqs.~(\ref{e:gtmd_1})--(\ref{e:gtmd_3}) are omitted. 
All GTMDs depend on the set of variables 
$(x,\xi,\vec{k}_T^2,\vec{k}_T \cdot \vec{\Delta}_T,\vec{\Delta}_T^2;\eta)$.

In the twist-3 case, characterized through a suppression by one power in $P^{+}$, 
we find
\begin{eqnarray}
 W_{\lambda \lambda'}^{[1]}
 &=& \frac{1}{2P^+} \, \bar{u}(p', \lambda') \, \bigg[
      E_{2,1}
      + \frac{i\sigma^{i+} k_T^i}{P^+} \, E_{2,2}
      + \frac{i\sigma^{i+} \Delta_T^i}{P^+} \, E_{2,3} \nonumber\\*
 & &  + \frac{i\sigma^{ij} k_T^i \Delta_T^j}{M^2} \, E_{2,4}
     \bigg] \, u(p, \lambda)
     \,, \label{e:gtmd_4}\\
 W_{\lambda \lambda'}^{[\gamma_5]}
 &=& \frac{1}{2P^+} \, \bar{u}(p', \lambda') \, \bigg[
      - \frac{i\varepsilon_T^{ij} k_T^i \Delta_T^j}{M^2} \, E_{2,5}
      + \frac{i\sigma^{i+}\gamma_5 k_T^i}{P^+} \, E_{2,6}
      + \frac{i\sigma^{i+}\gamma_5 \Delta_T^i}{P^+} \, E_{2,7} \nonumber\\*
 & &  + i\sigma^{+-}\gamma_5 \, E_{2,8}
     \bigg] \, u(p, \lambda)
     \,, \label{e:gtmd_5}\\
 W_{\lambda \lambda'}^{[\gamma^j]}
 &=& \frac{1}{2P^+} \, \bar{u}(p', \lambda') \, \bigg[
      \frac{k_T^j}{M} \, F_{2,1}
      + \frac{\Delta_T^j}{M} \, F_{2,2}
      + \frac{M \, i\sigma^{j+}}{P^+} \, F_{2,3} \nonumber\\*
 & &  + \frac{k_T^j \, i\sigma^{k+} k_T^k}{M \, P^+} \, F_{2,4}
      + \frac{\Delta_T^j \, i\sigma^{k+} k_T^k}{M \, P^+} \, F_{2,5}
      + \frac{\Delta_T^j \, i\sigma^{k+} \Delta_T^k}{M \, P^+} \, F_{2,6} \nonumber\\*
 & &  + \frac{i\sigma^{ij} k_T^i}{M} \, F_{2,7}
      + \frac{i\sigma^{ij} \Delta_T^i}{M} \, F_{2,8}
     \bigg] \, u(p, \lambda)
     \,, \label{e:gtmd_6}\\
 W_{\lambda \lambda'}^{[\gamma^j\gamma_5]}
 &=& \frac{1}{2P^+} \, \bar{u}(p', \lambda') \, \bigg[
      - \frac{i\varepsilon_T^{ij} k_T^i}{M} \, G_{2,1}
      - \frac{i\varepsilon_T^{ij} \Delta_T^i}{M} \, G_{2,2}
      + \frac{M \, i\sigma^{j+}\gamma_5}{P^+} \, G_{2,3} \nonumber\\*
 & &  + \frac{k_T^j \, i\sigma^{k+}\gamma_5 k_T^k}{M \, P^+} \, G_{2,4}
      + \frac{\Delta_T^j \, i\sigma^{k+}\gamma_5 k_T^k}{M \, P^+} \, G_{2,5}
      + \frac{\Delta_T^j \, i\sigma^{k+}\gamma_5 \Delta_T^k}{M \, P^+} \, G_{2,6} \nonumber\\*
 & &  + \frac{k_T^j \, i\sigma^{+-}\gamma_5}{M} \, G_{2,7}
      + \frac{\Delta_T^j \, i\sigma^{+-}\gamma_5}{M} \, G_{2,8}
     \bigg] \, u(p, \lambda)
     \,, \label{e:gtmd_7}\\
 W_{\lambda \lambda'}^{[i\sigma^{ij}\gamma_5]}
 &=& - \frac{i\varepsilon_T^{ij}}{2P^+} \, \bar{u}(p', \lambda') \, \bigg[
      H_{2,1}
      + \frac{i\sigma^{k+} k_T^k}{P^+} \, H_{2,2}
      + \frac{i\sigma^{k+} \Delta_T^k}{P^+} \, H_{2,3} \nonumber\\*
 & &  + \frac{i\sigma^{kl} k_T^k \Delta_T^l}{M^2} \, H_{2,4}
     \bigg] \, u(p, \lambda)
     \,, \label{e:gtmd_8}\\
 W_{\lambda \lambda'}^{[i\sigma^{+-}\gamma_5]}
 &=& \frac{1}{2P^+} \, \bar{u}(p', \lambda') \, \bigg[
      - \frac{i\varepsilon_T^{ij} k_T^i \Delta_T^j}{M^2} \, H_{2,5}
      + \frac{i\sigma^{i+}\gamma_5 k_T^i}{P^+} \, H_{2,6}
      + \frac{i\sigma^{i+}\gamma_5 \Delta_T^i}{P^+} \, H_{2,7} \nonumber\\*
 & &  + i\sigma^{+-}\gamma_5 \, H_{2,8}
     \bigg] \, u(p, \lambda)
     \,. \label{e:gtmd_9}
\end{eqnarray}
The twist-4 result, which is basically a copy of the twist-2 case, reads
\begin{eqnarray}
 W_{\lambda \lambda'}^{[\gamma^-]}
 &=& \frac{M}{2(P^+)^2} \, \bar{u}(p', \lambda') \, \bigg[
      F_{3,1}
      + \frac{i\sigma^{i+} k_T^i}{P^+} \, F_{3,2}
      + \frac{i\sigma^{i+} \Delta_T^i}{P^+} \, F_{3,3} \nonumber\\*
 & &  + \frac{i\sigma^{ij} k_T^i \Delta_T^j}{M^2} \, F_{3,4}
     \bigg] \, u(p, \lambda)
     \,, \label{e:gtmd_10}\\
 W_{\lambda \lambda'}^{[\gamma^-\gamma_5]}
 &=& \frac{M}{2(P^+)^2} \, \bar{u}(p', \lambda') \, \bigg[
      - \frac{i\varepsilon_T^{ij} k_T^i \Delta_T^j}{M^2} \, G_{3,1}
      + \frac{i\sigma^{i+}\gamma_5 k_T^i}{P^+} \, G_{3,2}
      + \frac{i\sigma^{i+}\gamma_5 \Delta_T^i}{P^+} \, G_{3,3} \nonumber\\*
 & &  + i\sigma^{+-} \, G_{3,4}
     \bigg] \, u(p, \lambda)
     \,, \label{e:gtmd_11}\\
 W_{\lambda \lambda'}^{[i\sigma^{j-}\gamma_5]}
 &=& \frac{M}{2(P^+)^2} \, \bar{u}(p', \lambda') \, \bigg[
      - \frac{i\varepsilon_T^{ij} k_T^i}{M} \, H_{3,1}
      - \frac{i\varepsilon_T^{ij} \Delta_T^i}{M} \, H_{3,2}
      + \frac{M \, i\sigma^{j+}\gamma_5}{P^+} \, H_{3,3} \nonumber\\*
 & &  + \frac{k_T^j \, i\sigma^{k+}\gamma_5 k_T^k}{M \, P^+} \, H_{3,4}
      + \frac{\Delta_T^j \, i\sigma^{k+}\gamma_5 k_T^k}{M \, P^+} \, H_{3,5}
      + \frac{\Delta_T^j \, i\sigma^{k+}\gamma_5 \Delta_T^k}{M \, P^+} \, H_{3,6} \nonumber\\*
 & &  + \frac{k_T^j \, i\sigma^{+-}\gamma_5}{M} \, H_{3,7}
      + \frac{\Delta_T^j \, i\sigma^{+-}\gamma_5}{M} \, H_{3,8}
     \bigg] \, u(p, \lambda)
     \,. \label{e:gtmd_12}
\end{eqnarray}
The twist-4 case is of course at most of academic interest but is included for 
completeness. 

%
%
%
\subsection{Properties}
Like in the case of the GPCFs we also consider the implications of hermiticity and 
time reversal on the GTMDs. 
Hermiticity leads to
\begin{equation}
 X^*(x,\xi,\vec{k}_T^2,\vec{k}_T \cdot \vec{\Delta}_T,\vec{\Delta}_T^2;\eta)
 =\pm X(x,-\xi,\vec{k}_T^2,-\vec{k}_T \cdot \vec{\Delta}_T,\vec{\Delta}_T^2;\eta) \,,
 \label{e:gtmd_hermiticity}
\end{equation}
with a plus sign for $X = E_{2,1}$, $E_{2,3}$, $E_{2,4}$, $E_{2,7}$, $F_{1,1}$, 
$F_{1,3}$, $F_{1,4}$, $F_{2,1}$, $F_{2,5}$, $F_{2,8}$, $F_{3,1}$, $F_{3,3}$, $F_{3,4}$, 
$G_{1,1}$, $G_{1,2}$, $G_{1,4}$, $G_{2,2}$, $G_{2,3}$, $G_{2,4}$, $G_{2,6}$, $G_{2,7}$, 
$G_{3,1}$, $G_{3,2}$, $G_{3,4}$, $H_{1,2}$, $H_{1,3}$, $H_{1,4}$, $H_{1,6}$, $H_{1,7}$, 
$H_{2,2}$, $H_{2,5}$, $H_{2,6}$, $H_{2,8}$, $H_{3,2}$, $H_{3,3}$, $H_{3,4}$, $H_{3,6}$, 
$H_{3,7}$ and a minus sign for all the other GTMDs. 
These results are a direct consequence of~(\ref{e:gpcf_hermiticity}) and the relations 
between GTMDs and GPCFs (see appendix~\ref{c:app_gtmd_gpcf} for the explicit formulas 
for twist-2). 
On the basis of~(\ref{e:gpcf_timereversal}) one obtains from time reversal
\begin{equation}
 X^*(x,\xi,\vec{k}_T^2,\vec{k}_T \cdot \vec{\Delta}_T,\vec{\Delta}_T^2;\eta) = 
 X(x,\xi,\vec{k}_T^2,\vec{k}_T \cdot \vec{\Delta}_T,\vec{\Delta}_T^2;-\eta) 
 \label{linear}
\end{equation}
for all GTMDs $X$. 
This means, in particular, that we can carry over eqs.~(\ref{e:gpcf_decomp}) 
and (\ref{e:gpcf_sign}) to the GTMD case and write
\begin{equation}
 X(x,\xi,\vec{k}_T^2,\vec{k}_T \cdot \vec{\Delta}_T,\vec{\Delta}_T^2;\eta)
 = X^{e}(x,\xi,\vec{k}_T^2,\vec{k}_T \cdot \vec{\Delta}_T,\vec{\Delta}_T^2) 
   +  i \, X^{o}(x,\xi,\vec{k}_T^2,\vec{k}_T \cdot \vec{\Delta}_T,\vec{\Delta}_T^2;\eta) \,, 
\end{equation}
with the real valued functions $X^{e}$ and $X^{o}$ respectively representing the  
real and imaginary part of the GTMD $X$. 
Only the T-odd part $X^{o}$ depends on the sign of $\eta$ according to 
\begin{equation}
 X^{o}(x,\xi,\vec{k}_T^2,\vec{k}_T \cdot \vec{\Delta}_T,\vec{\Delta}_T^2;\eta) = - 
 X^{o}(x,\xi,\vec{k}_T^2,\vec{k}_T \cdot \vec{\Delta}_T,\vec{\Delta}_T^2;-\eta) \,, 
\end{equation}
i.e., the imaginary parts of GTMDs defined with future-pointing and past-pointing 
Wilson lines have a reversed sign.

In order to give an estimate we have calculated the leading twist GTMDs in the 
scalar diquark model of the nucleon. 
The results are presented in appendix~\ref{c:app_gtmd_model}. 
Our treatment is restricted to lowest order in perturbation theory. 
To this order all T-odd parts of the GTMDs vanish --- a feature which is also 
well-known from spectator model calculations of T-odd TMDs. 
All the results listed in eqs.~(\ref{e:gtmd_model_1})--(\ref{e:gtmd_model_16}) are 
in accordance with the hermiticity constraint~(\ref{e:gtmd_hermiticity}).

%
%
%
\section{Projection of GTMDs onto TMDs and GPDs}\label{c:sec4}
In this section we consider the generalized $k_T$-dependent correlator in 
eq.~(\ref{e:corr_gtmd}) for the specific TMD-kinematics and the GPD-kinematics. 
This procedure provides the relations between the {\it mother distributions} (GTMDs) 
on the one hand and the TMDs as well as the GPDs on the other. 
On the basis of these results one can check whether there exists model-independent 
support for possible nontrivial relations between GPDs and TMDs.

%
%
%
\subsection{TMD-limit}
We start with the TMD-limit corresponding to a vanishing momentum transfer $\Delta = 0$. 
In this limit exactly half of the real-valued distributions vanish because they are 
odd as function of $\Delta$ due to the hermiticity constraint~(\ref{e:gtmd_hermiticity}): 
$E_{2,1}^o$, $E_{2,2}^e$, $E_{2,3}^o$, $E_{2,4}^o$, $E_{2,5}^e$, $E_{2,6}^e$, $E_{2,7}^o$, 
$E_{2,8}^e$, $F_{1,1}^o$, $F_{1,2}^e$, $F_{1,3}^o$, $F_{1,4}^o$, $F_{2,1}^o$, $F_{2,2}^e$, 
$F_{2,3}^e$, $F_{2,4}^e$, $F_{2,5}^o$, $F_{2,6}^e$, $F_{2,7}^e$, $F_{2,8}^o$, $F_{3,1}^o$, 
$F_{3,2}^e$, $F_{3,3}^o$, $F_{3,4}^o$, $G_{1,1}^o$, $G_{1,2}^o$, $G_{1,3}^e$, $G_{1,4}^o$, 
$G_{2,1}^e$, $G_{2,2}^o$, $G_{2,3}^o$, $G_{2,4}^o$, $G_{2,5}^e$, $G_{2,6}^o$, $G_{2,7}^o$, 
$G_{2,8}^e$, $G_{3,1}^o$, $G_{3,2}^o$, $G_{3,3}^e$, $G_{3,4}^o$, $H_{1,1}^e$, $H_{1,2}^o$, 
$H_{1,3}^o$, $H_{1,4}^o$, $H_{1,5}^e$, $H_{1,6}^o$, $H_{1,7}^o$, $H_{1,8}^e$, $H_{2,1}^e$, 
$H_{2,2}^o$, $H_{2,3}^e$, $H_{2,4}^e$, $H_{2,5}^o$, $H_{2,6}^o$, $H_{2,7}^e$, $H_{2,8}^o$, 
$H_{3,1}^e$, $H_{3,2}^o$, $H_{3,3}^o$, $H_{3,4}^o$, $H_{3,5}^e$, $H_{3,6}^o$, $H_{3,7}^o$, 
$H_{3,8}^e$. 
In addition, the distributions $E_{2,3}^e$, $E_{2,4}^e$, $E_{2,5}^o$, $E_{2,7}^e$, 
$F_{1,3}^e$, $F_{1,4}^e$, $F_{2,2}^o$, $F_{2,5}^e$, $F_{2,6}^o$, $F_{2,8}^e$, $F_{3,3}^e$, 
$F_{3,4}^e$, $G_{1,1}^e$, $G_{1,3}^o$, $G_{2,2}^e$, $G_{2,5}^o$, $G_{2,6}^e$, $G_{2,8}^o$, 
$G_{3,1}^e$, $G_{3,3}^o$, $H_{1,2}^e$, $H_{1,5}^o$, $H_{1,6}^e$, $H_{1,8}^o$, $H_{2,3}^o$, 
$H_{2,4}^o$, $H_{2,5}^e$, $H_{2,7}^o$, $H_{3,2}^e$, $H_{3,5}^o$, $H_{3,6}^e$, $H_{3,8}^o$ 
do not appear in the correlator any more, because they are multiplied by a coefficient which 
is linear in $\Delta$. 
Therefore, in the TMD-limit only the following 32 (20 T-even and 12 T-odd) distributions 
survive: $E_{2,1}^e$, $E_{2,2}^o$, $E_{2,6}^o$, $E_{2,8}^o$, $F_{1,1}^e$, $F_{1,2}^o$, 
$F_{2,1}^e$, $F_{2,3}^o$, $F_{2,4}^o$, $F_{2,7}^o$, $F_{3,1}^e$, $F_{3,2}^o$, $G_{1,2}^e$, 
$G_{1,4}^e$, $G_{2,1}^o$, $G_{2,3}^e$, $G_{2,4}^e$, $G_{2,7}^e$, $G_{3,2}^e$, $G_{3,4}^e$, 
$H_{1,1}^o$, $H_{1,3}^e$, $H_{1,4}^e$, $H_{1,7}^e$, $H_{2,1}^o$, $H_{2,2}^e$, $H_{2,6}^e$, 
$H_{2,8}^e$, $H_{3,1}^o$, $H_{3,3}^e$, $H_{3,4}^e$, $H_{3,7}^e$. 

The complete list of TMDs for a spin-1/2 hadron has been given in ref.~\cite{Goeke:2005hb} 
(see also the review article~\cite{Bacchetta:2006tn}). 
Here the spin vector 
\begin{equation}
S = \bigg[\, \lambda \frac{P^+}{M} \, , \, 
           - \lambda \frac{M}{2P^+} \, , \,
             \vec{S}_T \, \bigg]
\end{equation}
of the nucleon was introduced leading to the linear combination~\cite{Meissner:2007rx}
\begin{eqnarray}
 \Phi^{[\Gamma]}(P, x, \vec{k}_T, N; S; \eta)
 &=& \tfrac{1 + \lambda}{2} \, \Phi_{++}^{[\Gamma]}(P, x, \vec{k}_T, N; \eta)
     + \tfrac{1 - \lambda}{2} \, \Phi_{--}^{[\Gamma]}(P, x, \vec{k}_T, N; \eta) \nonumber\\*
 & & + \tfrac{S_T^1 - i S_T^2}{2} \, \Phi_{+-}^{[\Gamma]}(P, x, \vec{k}_T, N; \eta)
     + \tfrac{S_T^1 + i S_T^2}{2} \, \Phi_{-+}^{[\Gamma]}(P, x, \vec{k}_T, N; \eta) \,. \qquad
\end{eqnarray}
Now using the conventions of~\cite{Bacchetta:2006tn} for the TMDs one finds
the following explicit relations between the TMDs and the GTMDs:
\begin{eqnarray}
 f_1(x,\vec{k}_T^2) & = & F_{1,1}^e(x,0,\vec{k}_T^2,0,0) \,, \label{e:tmd_gtmd_1} \\
 f_{1T}^\bot(x,\vec{k}_T^2;\eta) & = & - F_{1,2}^o(x,0,\vec{k}_T^2,0,0;\eta) \,, \label{e:tmd_gtmd_2} \\
 g_{1L}(x,\vec{k}_T^2) & = & G_{1,4}^e(x,0,\vec{k}_T^2,0,0) \,, \label{e:tmd_gtmd_3} \\
 g_{1T}(x,\vec{k}_T^2) & = & G_{1,2}^e(x,0,\vec{k}_T^2,0,0) \,, \label{e:tmd_gtmd_4} \\
 h_1^\bot(x,\vec{k}_T^2;\eta) & = & - H_{1,1}^o(x,0,\vec{k}_T^2,0,0;\eta) \,, \label{e:tmd_gtmd_5} \\
 h_{1L}^\bot(x,\vec{k}_T^2) & = & H_{1,7}^e(x,0,\vec{k}_T^2,0,0) \,, \label{e:tmd_gtmd_6} \\
 h_{1T}(x,\vec{k}_T^2) & = & H_{1,3}^e(x,0,\vec{k}_T^2,0,0) \,, \label{e:tmd_gtmd_7} \\
 h_{1T}^\bot(x,\vec{k}_T^2) & = & H_{1,4}^e(x,0,\vec{k}_T^2,0,0) \,, \label{e:tmd_gtmd_8} \\
 e(x,\vec{k}_T^2) & = & E_{2,1}^e(x,0,\vec{k}_T^2,0,0) \,, \label{e:tmd_gtmd_9} \\
 e_L(x,\vec{k}_T^2;\eta) & = & - E_{2,8}^o(x,0,\vec{k}_T^2,0,0;\eta) \,, \label{e:tmd_gtmd_10} \\
 e_T(x,\vec{k}_T^2;\eta) & = & - E_{2,6}^o(x,0,\vec{k}_T^2,0,0;\eta) \,, \label{e:tmd_gtmd_11} \\
 e_T^\bot(x,\vec{k}_T^2;\eta) & = & - E_{2,2}^o(x,0,\vec{k}_T^2,0,0;\eta) \,, \label{e:tmd_gtmd_12} \\
 f^\bot(x,\vec{k}_T^2) & = & F_{2,1}^e(x,0,\vec{k}_T^2,0,0) \,, \label{e:tmd_gtmd_13} \\
 f_L^\bot(x,\vec{k}_T^2;\eta) & = & F_{2,7}^o(x,0,\vec{k}_T^2,0,0;\eta) \,, \label{e:tmd_gtmd_14} \\
 f_T'(x,\vec{k}_T^2;\eta) & = & F_{2,3}^o(x,0,\vec{k}_T^2,0,0;\eta) \,, \label{e:tmd_gtmd_15} \\
 f_T^\bot(x,\vec{k}_T^2;\eta) & = & F_{2,4}^o(x,0,\vec{k}_T^2,0,0;\eta) \,, \label{e:tmd_gtmd_16} \\
 g^\bot(x,\vec{k}_T^2;\eta) & = & - G_{2,1}^o(x,0,\vec{k}_T^2,0,0;\eta) \,, \label{e:tmd_gtmd_17} \\
 g_L^\bot(x,\vec{k}_T^2) & = & G_{2,7}^e(x,0,\vec{k}_T^2,0,0) \,, \label{e:tmd_gtmd_18} \\
 g_T'(x,\vec{k}_T^2) & = & G_{2,3}^e(x,0,\vec{k}_T^2,0,0) \,, \label{e:tmd_gtmd_19} \\
 g_T^\bot(x,\vec{k}_T^2) & = & G_{2,4}^e(x,0,\vec{k}_T^2,0,0) \,, \label{e:tmd_gtmd_20} \\
 h(x,\vec{k}_T^2;\eta) & = & - H_{2,1}^o(x,0,\vec{k}_T^2,0,0;\eta) \,, \label{e:tmd_gtmd_21}\\
 h_L(x,\vec{k}_T^2) & = & H_{2,8}^e(x,0,\vec{k}_T^2,0,0) \,, \label{e:tmd_gtmd_22} \\
 h_T(x,\vec{k}_T^2) & = & H_{2,6}^e(x,0,\vec{k}_T^2,0,0) \,, \label{e:tmd_gtmd_23} \\
 h_T^\bot(x,\vec{k}_T^2) & = & H_{2,2}^e(x,0,\vec{k}_T^2,0,0) \,, \label{e:tmd_gtmd_24} \\
 f_3(x,\vec{k}_T^2) & = & F_{3,1}^e(x,0,\vec{k}_T^2,0,0) \,, \label{e:tmd_gtmd_25} \\
 f_{3T}^\bot(x,\vec{k}_T^2;\eta) & = & - F_{3,2}^o(x,0,\vec{k}_T^2,0,0;\eta) \,, \label{e:tmd_gtmd_26} \\
 g_{3L}(x,\vec{k}_T^2) & = & G_{3,4}^e(x,0,\vec{k}_T^2,0,0) \,, \label{e:tmd_gtmd_27} \\
 g_{3T}(x,\vec{k}_T^2) & = & G_{3,2}^e(x,0,\vec{k}_T^2,0,0) \,, \label{e:tmd_gtmd_28} \\
 h_3^\bot(x,\vec{k}_T^2;\eta) & = & - H_{3,1}^o(x,0,\vec{k}_T^2,0,0;\eta) \,, \label{e:tmd_gtmd_29} \\
 h_{3L}^\bot(x,\vec{k}_T^2) & = & H_{3,7}^e(x,0,\vec{k}_T^2,0,0) \,, \label{e:tmd_gtmd_30} \\
 h_{3T}(x,\vec{k}_T^2) & = & H_{3,3}^e(x,0,\vec{k}_T^2,0,0) \,, \label{e:tmd_gtmd_31} \\
 h_{3T}^\bot(x,\vec{k}_T^2) & = & H_{3,4}^e(x,0,\vec{k}_T^2,0,0) \,. \label{e:tmd_gtmd_32}
\end{eqnarray}
These results are obtained by means of eqs.~(\ref{e:corr_tmd}) 
and~(\ref{e:gtmd_1})--(\ref{e:gtmd_12}). 
The 12 TMDs $f_{1T}^\bot$, $h_1^\bot$, $e_L$, $e_T$, $e_T^\bot$, $f_L^\bot$, $f_T$, 
$f_T^\bot$, $g^\bot$, $h$, $f_{3T}^\bot$, $h_3^\bot$ are T-odd and are related to 
T-odd parts of GTMDs.

%
%
%
\subsection{GPD-limit}
In a second step we focus on the GPD-limit which appears when integrating 
upon the transverse parton momentum $\vec{k}_T$. 
As already discussed after eq.~(\ref{e:corr_gpd}) the dependence on $\eta$ drops 
out in this case which implies, in particular, that all effects of T-odd parts of 
GTMDs disappear. 
In the literature only the twist-2 and the chiral-even twist-3 GPDs have been 
introduced~\cite{Diehl:2001pm,Kiptily:2002nx}. 
Therefore, we give here for the first time a complete list of GPDs for all twists. 
The GPDs parameterize the correlator in~(\ref{e:corr_gpd}). 
One finds 8 GPDs for twist-2, 16 GPDs for twist-3, and 8 GPDs for twist-4.

To be explicit the GPDs can be defined according to
\begin{eqnarray}
 F_{\lambda \lambda'}^{[\gamma^+]}
 &=& \frac{1}{2P^+} \, \bar{u}(p', \lambda') \, \bigg[
      \gamma^+ \, H(x, \xi, t)
      + \frac{i\sigma^{+\Delta}}{2M} \, E(x, \xi, t)
     \bigg] \, u(p, \lambda) \,,
     \label{e:gpd_1}\\
 F_{\lambda \lambda'}^{[\gamma^+\gamma_5]}
 &=& \frac{1}{2P^+} \, \bar{u}(p', \lambda') \, \bigg[
      \gamma^+\gamma_5 \, \tilde{H}(x, \xi, t)
      + \frac{\Delta^+ \gamma_5}{2M} \, \tilde{E}(x, \xi, t)
     \bigg] \, u(p, \lambda) \,,
     \label{e:gpd_2}\\
 F_{\lambda \lambda'}^{[i\sigma^{j+}\gamma_5]}
 &=& - \frac{i\varepsilon_T^{ij}}{2P^+} \, \bar{u}(p', \lambda') \, \bigg[
      i\sigma^{+i} \, H_T(x, \xi, t)
      + \frac{\gamma^+ \Delta_T^i  - \Delta^+ \gamma^i}{2M} \, E_T(x, \xi, t) \nonumber\\*
 & & + \frac{P^+ \Delta_T^i  - \Delta^+ P_T^i}{M^2} \, \tilde{H}_T(x, \xi, t)
      + \frac{\gamma^+ P_T^i  - P^+ \gamma^i}{M} \, \tilde{E}_T(x, \xi, t)
     \bigg] \, u(p, \lambda) \,,
 \label{e:gpd_3}\\
 F_{\lambda \lambda'}^{[1]}
 &=& \frac{M}{2(P^+)^2} \, \bar{u}(p', \lambda') \, \bigg[
      \gamma^+ \, H_2(x, \xi, t)
      + \frac{i\sigma^{+\Delta}}{2M} \, E_2(x, \xi, t)
     \bigg] \, u(p, \lambda) \,,
     \label{e:gpd_4}\\
 F_{\lambda \lambda'}^{[\gamma_5]}
 &=& \frac{M}{2(P^+)^2} \, \bar{u}(p', \lambda') \, \bigg[
      \gamma^+\gamma_5 \, \tilde{H}_2(x, \xi, t)
      + \frac{P^+ \gamma_5}{M} \, \tilde{E}_2(x, \xi, t)
     \bigg] \, u(p, \lambda) \,,
     \label{e:gpd_5}\\
 F_{\lambda \lambda'}^{[\gamma^j]}
 &=& \frac{M}{2(P^+)^2} \, \bar{u}(p', \lambda') \, \bigg[
      i\sigma^{+j} \, H_{2T}(x, \xi, t)
      + \frac{\gamma^+ \Delta_T^j  - \Delta^+ \gamma^j}{2M} \, E_{2T}(x, \xi, t) \nonumber\\*
 & & + \frac{P^+ \Delta_T^j  - \Delta^+ P_T^j}{M^2} \, \tilde{H}_{2T}(x, \xi, t)
      + \frac{\gamma^+ P_T^j  - P^+ \gamma^j}{M} \, \tilde{E}_{2T}(x, \xi, t)
     \bigg] \, u(p, \lambda) \,,
 \label{e:gpd_6}\\
 F_{\lambda \lambda'}^{[\gamma^j\gamma_5]}
 &=& - \frac{i\varepsilon_T^{ij} M}{2(P^+)^2} \, \bar{u}(p', \lambda') \, \bigg[
      i\sigma^{+i} \, H'_{2T}(x, \xi, t)
      + \frac{\gamma^+ \Delta_T^i  - \Delta^+ \gamma^i}{2M} \, E'_{2T}(x, \xi, t) \nonumber\\*
 & & + \frac{P^+ \Delta_T^i  - \Delta^+ P_T^i}{M^2} \, \tilde{H}'_{2T}(x, \xi, t)
      + \frac{\gamma^+ P_T^i  - P^+ \gamma^i}{M} \, \tilde{E}'_{2T}(x, \xi, t)
     \bigg] \, u(p, \lambda) \,,
 \label{e:gpd_7}\\
 F_{\lambda \lambda'}^{[i\sigma^{ij}\gamma_5]}
 &=& - \frac{i\varepsilon_T^{ij} M}{2(P^+)^2} \, \bar{u}(p', \lambda') \, \bigg[
      \gamma^+ \, H'_2(x, \xi, t)
      + \frac{i\sigma^{+\Delta}}{2M} \, E'_2(x, \xi, t)
     \bigg] \, u(p, \lambda) \,,
     \label{e:gpd_8}\\
 F_{\lambda \lambda'}^{[i\sigma^{+-}\gamma_5]}
 &=& \frac{M}{2(P^+)^2} \, \bar{u}(p', \lambda') \, \bigg[
      \gamma^+\gamma_5 \, \tilde{H}'_2(x, \xi, t)
      + \frac{P^+ \gamma_5}{M} \, \tilde{E}'_2(x, \xi, t)
     \bigg] \, u(p, \lambda) \,,
     \label{e:gpd_9}\\
 F_{\lambda \lambda'}^{[\gamma^-]}
 &=& \frac{M^2}{2(P^+)^3} \, \bar{u}(p', \lambda') \, \bigg[
      \gamma^+ \, H_3(x, \xi, t)
      + \frac{i\sigma^{+\Delta}}{2M} \, E_3(x, \xi, t)
     \bigg] \, u(p, \lambda) \,,
     \label{e:gpd_10}\\
 F_{\lambda \lambda'}^{[\gamma^-\gamma_5]}
 &=& \frac{M^2}{2(P^+)^3} \, \bar{u}(p', \lambda') \, \bigg[
      \gamma^+\gamma_5 \, \tilde{H}_3(x, \xi, t)
      + \frac{\Delta^+ \gamma_5}{2M} \, \tilde{E}_3(x, \xi, t)
     \bigg] \, u(p, \lambda) \,,
     \label{e:gpd_11}\\
 F_{\lambda \lambda'}^{[i\sigma^{j-}\gamma_5]}
 &=& - \frac{i\varepsilon_T^{ij} M^2}{2(P^+)^3} \, \bar{u}(p', \lambda') \, \bigg[
      i\sigma^{+i} \, H_{3T}(x, \xi, t)
      + \frac{\gamma^+ \Delta_T^i  - \Delta^+ \gamma^i}{2M} \, E_{3T}(x, \xi, t) \nonumber\\*
 & & + \frac{P^+ \Delta_T^i  - \Delta^+ P_T^i}{M^2} \, \tilde{H}_{3T}(x, \xi, t)
      + \frac{\gamma^+ P_T^i  - P^+ \gamma^i}{M} \, \tilde{E}_{3T}(x, \xi, t)
     \bigg] \, u(p, \lambda) \,, \quad
 \label{e:gpd_12}
\end{eqnarray}
where $t = \Delta^2$. 
The structure of the traces in~(\ref{e:gpd_1})--(\ref{e:gpd_12}) follows readily  
from eqs.~(\ref{e:gtmd_1})--(\ref{e:gtmd_12}) if one keeps in mind that after 
integrating upon $\vec{k}_T$ the only transverse vector left is $\vec{\Delta}_T$. 
Altogether there exist 32 GPDs corresponding to the number of TMDs. 
The 16 GPDs $H$, $E$, $\tilde{H}$, $\tilde{E}$, $H_{2T}$, $E_{2T}$, 
$\tilde{H}_{2T}$, $\tilde{E}_{2T}$, $H'_{2T}$, $E'_{2T}$, $\tilde{H}'_{2T}$, 
$\tilde{E}'_{2T}$, $H_3$, $E_3$, $\tilde{H}_3$, $\tilde{E}_3$ are chiral-even, 
while the remaining ones are chiral-odd. 
The definition of the twist-2 GPDs corresponds follows the common 
definition \cite{Diehl:2001pm}. 
The chiral-even twist-3 GPDs $H_{2T}$, $E_{2T}$, $\tilde{H}_{2T}$, $\tilde{E}_{2T}$, 
$H'_{2T}$, $E'_{2T}$, $\tilde{H}'_{2T}$, $\tilde{E}'_{2T}$ are related to the 
functions $G_1$, $G_2$, $G_3$, $G_4$, $\tilde{G}_1$, $\tilde{G}_2$, $\tilde{G}_3$, 
$\tilde{G}_4$ that were introduced in ref.~\cite{Kiptily:2002nx}. 

It is now straightforward to write down the following expressions for the GPDs in 
terms of $k_T$-integrals of GTMDs:
\begin{eqnarray} 
 H(x,\xi,t) & = & \int d^2\vec{k}_T \, \bigg[
  F_{1,1}^e
  + 2 \xi^2 \bigg(
   \frac{\vec{k}_T \cdot \vec{\Delta}_T}{\vec{\Delta}_T^2} \, F_{1,2}^e
   + F_{1,3}^e
  \bigg)
 \bigg] \,, \label{e:gpd_gtmd_1} \\
 E(x,\xi,t) & = & \int d^2\vec{k}_T \, \bigg[
  - F_{1,1}^e
  + 2 (1 - \xi^2) \bigg(
   \frac{\vec{k}_T \cdot \vec{\Delta}_T}{\vec{\Delta}_T^2} \, F_{1,2}^e
   + F_{1,3}^e
  \bigg)
 \bigg] \,, \label{e:gpd_gtmd_2} \\
 \tilde{H}(x,\xi,t) & = & \int d^2\vec{k}_T \, \bigg[
  2 \xi \bigg(
   \frac{\vec{k}_T \cdot \vec{\Delta}_T}{\vec{\Delta}_T^2} \, G_{1,2}^e
   + G_{1,3}^e
  \bigg)
  + G_{1,4}^e \bigg] \,, \label{e:gpd_gtmd_3} \\
 \tilde{E}(x,\xi,t) & = & \int d^2\vec{k}_T \, \bigg[
  \frac{2(1 - \xi^2)}{\xi} \bigg(
   \frac{\vec{k}_T \cdot \vec{\Delta}_T}{\vec{\Delta}_T^2} \, G_{1,2}^e
   + G_{1,3}^e
  \bigg)
  - G_{1,4}^e
 \bigg] \,, \label{e:gpd_gtmd_4} \\
 H_T(x,\xi,t) & = & \int d^2\vec{k}_T \, \bigg[
  H_{1,3}^e
  + \frac{\vec{\Delta}_T^2}{M^2} \bigg(
   \frac{(\vec{k}_T \cdot \vec{\Delta}_T)^2}{(\vec{\Delta}_T^2)^2} \, H_{1,4}^e
   + \frac{\vec{k}_T \cdot \vec{\Delta}_T}{\vec{\Delta}_T^2} \, H_{1,5}^e
   + H_{1,6}^e
  \bigg) \nonumber\\* && \qquad\qquad
  - \frac{\xi(\vec{\Delta}_T^2 + 4M^2)}{2(1 - \xi^2)M^2} \bigg(
   \frac{\vec{k}_T \cdot \vec{\Delta}_T}{\vec{\Delta}_T^2} \, H_{1,7}^e
   + H_{1,8}^e
  \bigg)
 \bigg] \,, \label{e:gpd_gtmd_5} \\
 E_T(x,\xi,t) & = & \int d^2\vec{k}_T \, \bigg[
  4 \bigg(
   \frac{2 (\vec{k}_T \cdot \vec{\Delta}_T)^2 - \vec{k}_T^2 \vec{\Delta}_T^2}{(\vec{\Delta}_T^2)^2} \, H_{1,4}^e
   + \frac{\vec{k}_T \cdot \vec{\Delta}_T}{\vec{\Delta}_T^2} \, H_{1,5}^e
   + H_{1,6}^e
  \bigg)
 \bigg] \,, \label{e:gpd_gtmd_6} \\
 \tilde{H}_T(x,\xi,t) & = & \int d^2\vec{k}_T \, \bigg[
  \bigg(
   \frac{\vec{k}_T \cdot \vec{\Delta}_T}{\vec{\Delta}_T^2} \, H_{1,1}^e
   + H_{1,2}^e
  \bigg) \nonumber\\* && \qquad\qquad
  - 2(1 - \xi^2) \bigg(
   \frac{2 (\vec{k}_T \cdot \vec{\Delta}_T)^2 - \vec{k}_T^2 \vec{\Delta}_T^2}{(\vec{\Delta}_T^2)^2} \, H_{1,4}^e
   + \frac{\vec{k}_T \cdot \vec{\Delta}_T}{\vec{\Delta}_T^2} \, H_{1,5}^e
   + H_{1,6}^e
  \bigg) \nonumber\\* && \qquad\qquad
  + \xi \bigg(
   \frac{\vec{k}_T \cdot \vec{\Delta}_T}{\vec{\Delta}_T^2} \, H_{1,7}^e
   + H_{1,8}^e
  \bigg)
 \bigg] \,, \label{e:gpd_gtmd_7} \\
 \tilde{E}_T(x,\xi,t) & = & \int d^2\vec{k}_T \, \bigg[
  4\xi \bigg(
   \frac{2 (\vec{k}_T \cdot \vec{\Delta}_T)^2 - \vec{k}_T^2 \vec{\Delta}_T^2}{(\vec{\Delta}_T^2)^2} \, H_{1,4}^e
   + \frac{\vec{k}_T \cdot \vec{\Delta}_T}{\vec{\Delta}_T^2} \, H_{1,5}^e
   + H_{1,6}^e
  \bigg) \nonumber\\* && \qquad\qquad
  + 2 \bigg(
   \frac{\vec{k}_T \cdot \vec{\Delta}_T}{\vec{\Delta}_T^2} \, H_{1,7}^e
   + H_{1,8}^e
  \bigg)
 \bigg] \,, \label{e:gpd_gtmd_8} \\
 H_2(x,\xi,t) & = & \int d^2\vec{k}_T \, \bigg[
  E_{2,1}^e
  + 2 \xi^2 \bigg(
   \frac{\vec{k}_T \cdot \vec{\Delta}_T}{\vec{\Delta}_T^2} \, E_{2,2}^e
   + E_{2,3}^e
  \bigg)
 \bigg] \,, \label{e:gpd_gtmd_9} \\
 E_2(x,\xi,t) & = & \int d^2\vec{k}_T \, \bigg[
  - E_{2,1}^e
  + 2 (1 - \xi^2) \bigg(
   \frac{\vec{k}_T \cdot \vec{\Delta}_T}{\vec{\Delta}_T^2} \, E_{2,2}^e
   + E_{2,3}^e
  \bigg)
 \bigg] \,, \label{e:gpd_gtmd_10} \\
 \tilde{H}_2(x,\xi,t) & = & \int d^2\vec{k}_T \, \bigg[
  2 \xi \bigg(
   \frac{\vec{k}_T \cdot \vec{\Delta}_T}{\vec{\Delta}_T^2} \, E_{2,6}^e
   + E_{2,7}^e
  \bigg)
  + E_{2,8}^e \bigg] \,, \label{e:gpd_gtmd_11} \\
 \tilde{E}_2(x,\xi,t) & = & \int d^2\vec{k}_T \, \bigg[
  - 2(1 - \xi^2) \bigg(
   \frac{\vec{k}_T \cdot \vec{\Delta}_T}{\vec{\Delta}_T^2} \, E_{2,6}^e
   + E_{2,7}^e
  \bigg)
  + \xi E_{2,8}^e
 \bigg] \,, \label{e:gpd_gtmd_12} \\
 H'_2(x,\xi,t) & = & \int d^2\vec{k}_T \, \bigg[
  H_{2,1}^e
  + 2 \xi^2 \bigg(
   \frac{\vec{k}_T \cdot \vec{\Delta}_T}{\vec{\Delta}_T^2} \, H_{2,2}^e
   + H_{2,3}^e
  \bigg)
 \bigg] \,, \label{e:gpd_gtmd_13} \\
 E'_2(x,\xi,t) & = & \int d^2\vec{k}_T \, \bigg[
  - H_{2,1}^e
  + 2 (1 - \xi^2) \bigg(
   \frac{\vec{k}_T \cdot \vec{\Delta}_T}{\vec{\Delta}_T^2} \, H_{2,2}^e
   + H_{2,3}^e
  \bigg)
 \bigg] \,, \label{e:gpd_gtmd_14} \\
 \tilde{H}'_2(x,\xi,t) & = & \int d^2\vec{k}_T \, \bigg[
  2 \xi \bigg(
   \frac{\vec{k}_T \cdot \vec{\Delta}_T}{\vec{\Delta}_T^2} \, H_{2,6}^e
   + H_{2,7}^e
  \bigg)
  + H_{2,8}^e \bigg] \,, \label{e:gpd_gtmd_15} \\
 \tilde{E}'_2(x,\xi,t) & = & \int d^2\vec{k}_T \, \bigg[
  - 2(1 - \xi^2) \bigg(
   \frac{\vec{k}_T \cdot \vec{\Delta}_T}{\vec{\Delta}_T^2} \, H_{2,6}^e
   + H_{2,7}^e
  \bigg)
  + \xi H_{2,8}^e
 \bigg] \,, \label{e:gpd_gtmd_16} \\
 H_{2T}(x,\xi,t) & = & \int d^2\vec{k}_T \, \bigg[
  -F_{2,3}^e
  + \frac{(\vec{k}_T \cdot \vec{\Delta}_T)^2 - \vec{k}_T^2 \vec{\Delta}_T^2}{M^2 \, \vec{\Delta}_T^2} \, F_{2,4}^e
  \nonumber\\* && \qquad\qquad
  + \frac{\xi(\vec{\Delta}_T^2 + 4M^2)}{2(1 - \xi^2)M^2} \bigg(
   \frac{\vec{k}_T \cdot \vec{\Delta}_T}{\vec{\Delta}_T^2} \, F_{2,7}^e
   + F_{2,8}^e
  \bigg)
 \bigg] \,, \label{e:gpd_gtmd_17} \\
 E_{2T}(x,\xi,t) & = & \int d^2\vec{k}_T \, \bigg[
  4 \bigg(
   \frac{2 (\vec{k}_T \cdot \vec{\Delta}_T)^2 - \vec{k}_T^2 \vec{\Delta}_T^2}{(\vec{\Delta}_T^2)^2} \, F_{2,4}^e
   + \frac{\vec{k}_T \cdot \vec{\Delta}_T}{\vec{\Delta}_T^2} \, F_{2,5}^e
   + F_{2,6}^e
  \bigg)
 \bigg] \,, \label{e:gpd_gtmd_18} \\
 \tilde{H}_{2T}(x,\xi,t) & = & \int d^2\vec{k}_T \, \bigg[
  \bigg(
   \frac{\vec{k}_T \cdot \vec{\Delta}_T}{\vec{\Delta}_T^2} \, F_{2,1}^e
   + F_{2,2}^e
  \bigg) \nonumber\\* && \qquad\qquad
  - 2(1 - \xi^2) \bigg(
   \frac{2 (\vec{k}_T \cdot \vec{\Delta}_T)^2 - \vec{k}_T^2 \vec{\Delta}_T^2}{(\vec{\Delta}_T^2)^2} \, F_{2,4}^e
   + \frac{\vec{k}_T \cdot \vec{\Delta}_T}{\vec{\Delta}_T^2} \, F_{2,5}^e
   + F_{2,6}^e
  \bigg) \nonumber\\* && \qquad\qquad
  - \xi \bigg(
   \frac{\vec{k}_T \cdot \vec{\Delta}_T}{\vec{\Delta}_T^2} \, F_{2,7}^e
   + F_{2,8}^e
  \bigg)
 \bigg] \,, \label{e:gpd_gtmd_19} \\
 \tilde{E}_{2T}(x,\xi,t) & = & \int d^2\vec{k}_T \, \bigg[
  4\xi \bigg(
   \frac{2 (\vec{k}_T \cdot \vec{\Delta}_T)^2 - \vec{k}_T^2 \vec{\Delta}_T^2}{(\vec{\Delta}_T^2)^2} \, F_{2,4}^e
   + \frac{\vec{k}_T \cdot \vec{\Delta}_T}{\vec{\Delta}_T^2} \, F_{2,5}^e
   + F_{2,6}^e
  \bigg) \nonumber\\* && \qquad\qquad
  - 2 \bigg(
   \frac{\vec{k}_T \cdot \vec{\Delta}_T}{\vec{\Delta}_T^2} \, F_{2,7}^e
   + F_{2,8}^e
  \bigg)
 \bigg] \,, \label{e:gpd_gtmd_20}\\
 H'_{2T}(x,\xi,t) & = & \int d^2\vec{k}_T \, \bigg[
  G_{2,3}^e
  + \frac{\vec{\Delta}_T^2}{M^2} \bigg(
   \frac{(\vec{k}_T \cdot \vec{\Delta}_T)^2}{(\vec{\Delta}_T^2)^2} \, G_{2,4}^e
   + \frac{\vec{k}_T \cdot \vec{\Delta}_T}{\vec{\Delta}_T^2} \, G_{2,5}^e
   + G_{2,6}^e
  \bigg) \nonumber\\* && \qquad\qquad
  - \frac{\xi(\vec{\Delta}_T^2 + 4M^2)}{2(1 - \xi^2)M^2} \bigg(
   \frac{\vec{k}_T \cdot \vec{\Delta}_T}{\vec{\Delta}_T^2} \, G_{2,7}^e
   + G_{2,8}^e
  \bigg)
 \bigg] \,, \label{e:gpd_gtmd_21} \\
 E'_{2T}(x,\xi,t) & = & \int d^2\vec{k}_T \, \bigg[
  4 \bigg(
   \frac{2 (\vec{k}_T \cdot \vec{\Delta}_T)^2 - \vec{k}_T^2 \vec{\Delta}_T^2}{(\vec{\Delta}_T^2)^2} \, G_{2,4}^e
   + \frac{\vec{k}_T \cdot \vec{\Delta}_T}{\vec{\Delta}_T^2} \, G_{2,5}^e
   + G_{2,6}^e
  \bigg)
 \bigg] \,, \label{e:gpd_gtmd_22} \\
 \tilde{H}'_{2T}(x,\xi,t) & = & \int d^2\vec{k}_T \, \bigg[
  \bigg(
   \frac{\vec{k}_T \cdot \vec{\Delta}_T}{\vec{\Delta}_T^2} \, G_{2,1}^e
   + G_{2,2}^e
  \bigg) \nonumber\\* && \qquad\qquad
  - 2(1 - \xi^2) \bigg(
   \frac{2 (\vec{k}_T \cdot \vec{\Delta}_T)^2 - \vec{k}_T^2 \vec{\Delta}_T^2}{(\vec{\Delta}_T^2)^2} \, G_{2,4}^e
   + \frac{\vec{k}_T \cdot \vec{\Delta}_T}{\vec{\Delta}_T^2} \, G_{2,5}^e
   + G_{2,6}^e
  \bigg) \nonumber\\* && \qquad\qquad
  + \xi \bigg(
   \frac{\vec{k}_T \cdot \vec{\Delta}_T}{\vec{\Delta}_T^2} \, G_{2,7}^e
   + G_{2,8}^e
  \bigg)
 \bigg] \,, \label{e:gpd_gtmd_23} \\
 \tilde{E}'_{2T}(x,\xi,t) & = & \int d^2\vec{k}_T \, \bigg[
  4\xi \bigg(
   \frac{2 (\vec{k}_T \cdot \vec{\Delta}_T)^2 - \vec{k}_T^2 \vec{\Delta}_T^2}{(\vec{\Delta}_T^2)^2} \, G_{2,4}^e
   + \frac{\vec{k}_T \cdot \vec{\Delta}_T}{\vec{\Delta}_T^2} \, G_{2,5}^e
   + G_{2,6}^e
  \bigg) \nonumber\\* && \qquad\qquad
  + 2 \bigg(
   \frac{\vec{k}_T \cdot \vec{\Delta}_T}{\vec{\Delta}_T^2} \, G_{2,7}^e
   + G_{2,8}^e
  \bigg)
 \bigg] \,, \label{e:gpd_gtmd_24}\\
 H_3(x,\xi,t) & = & \int d^2\vec{k}_T \, \bigg[
  F_{3,1}^e
  + 2 \xi^2 \bigg(
   \frac{\vec{k}_T \cdot \vec{\Delta}_T}{\vec{\Delta}_T^2} \, F_{3,2}^e
   + F_{3,3}^e
  \bigg)
 \bigg] \,, \label{e:gpd_gtmd_25} \\
 E_3(x,\xi,t) & = & \int d^2\vec{k}_T \, \bigg[
  - F_{3,1}^e
  + 2 (1 - \xi^2) \bigg(
   \frac{\vec{k}_T \cdot \vec{\Delta}_T}{\vec{\Delta}_T^2} \, F_{3,2}^e
   + F_{3,3}^e
  \bigg)
 \bigg] \,, \label{e:gpd_gtmd_26} \\
 \tilde{H}_3(x,\xi,t) & = & \int d^2\vec{k}_T \, \bigg[
  2 \xi \bigg(
   \frac{\vec{k}_T \cdot \vec{\Delta}_T}{\vec{\Delta}_T^2} \, G_{3,2}^e
   + G_{3,3}^e
  \bigg)
  + G_{3,4}^e \bigg] \,, \label{e:gpd_gtmd_27} \\
 \tilde{E}_3(x,\xi,t) & = & \int d^2\vec{k}_T \, \bigg[
  \frac{2(1 - \xi^2)}{\xi} \bigg(
   \frac{\vec{k}_T \cdot \vec{\Delta}_T}{\vec{\Delta}_T^2} \, G_{3,2}^e
   + G_{3,3}^e
  \bigg)
  - G_{3,4}^e
 \bigg] \,, \label{e:gpd_gtmd_28} \\
 H_{3T}(x,\xi,t) & = & \int d^2\vec{k}_T \, \bigg[
  H_{3,3}^e
  + \frac{\vec{\Delta}_T^2}{M^2} \bigg(
   \frac{(\vec{k}_T \cdot \vec{\Delta}_T)^2}{(\vec{\Delta}_T^2)^2} \, H_{3,4}^e
   + \frac{\vec{k}_T \cdot \vec{\Delta}_T}{\vec{\Delta}_T^2} \, H_{3,5}^e
   + H_{3,6}^e
  \bigg) \nonumber\\* && \qquad\qquad
  - \frac{\xi(\vec{\Delta}_T^2 + 4M^2)}{2(1 - \xi^2)M^2} \bigg(
   \frac{\vec{k}_T \cdot \vec{\Delta}_T}{\vec{\Delta}_T^2} \, H_{3,7}^e
   + H_{3,8}^e
  \bigg)
 \bigg] \,, \label{e:gpd_gtmd_29} \\
 E_{3T}(x,\xi,t) & = & \int d^2\vec{k}_T \, \bigg[
  4 \bigg(
   \frac{2 (\vec{k}_T \cdot \vec{\Delta}_T)^2 - \vec{k}_T^2 \vec{\Delta}_T^2}{(\vec{\Delta}_T^2)^2} \, H_{3,4}^e
   + \frac{\vec{k}_T \cdot \vec{\Delta}_T}{\vec{\Delta}_T^2} \, H_{3,5}^e
   + H_{3,6}^e
  \bigg)
 \bigg] \,, \label{e:gpd_gtmd_30} \\
 \tilde{H}_{3T}(x,\xi,t) & = & \int d^2\vec{k}_T \, \bigg[
  \bigg(
   \frac{\vec{k}_T \cdot \vec{\Delta}_T}{\vec{\Delta}_T^2} \, H_{3,1}^e
   + H_{3,2}^e
  \bigg) \nonumber\\* && \qquad\qquad
  - 2(1 - \xi^2) \bigg(
   \frac{2 (\vec{k}_T \cdot \vec{\Delta}_T)^2 - \vec{k}_T^2 \vec{\Delta}_T^2}{(\vec{\Delta}_T^2)^2} \, H_{3,4}^e
   + \frac{\vec{k}_T \cdot \vec{\Delta}_T}{\vec{\Delta}_T^2} \, H_{3,5}^e
   + H_{3,6}^e
  \bigg) \nonumber\\* && \qquad\qquad
  + \xi \bigg(
   \frac{\vec{k}_T \cdot \vec{\Delta}_T}{\vec{\Delta}_T^2} \, H_{3,7}^e
   + H_{3,8}^e
  \bigg)
 \bigg] \,, \label{e:gpd_gtmd_31} \\
 \tilde{E}_{3T}(x,\xi,t) & = & \int d^2\vec{k}_T \, \bigg[
  4\xi \bigg(
   \frac{2 (\vec{k}_T \cdot \vec{\Delta}_T)^2 - \vec{k}_T^2 \vec{\Delta}_T^2}{(\vec{\Delta}_T^2)^2} \, H_{3,4}^e
   + \frac{\vec{k}_T \cdot \vec{\Delta}_T}{\vec{\Delta}_T^2} \, H_{3,5}^e
   + H_{3,6}^e
  \bigg) \nonumber\\* && \qquad\qquad
  + 2 \bigg(
   \frac{\vec{k}_T \cdot \vec{\Delta}_T}{\vec{\Delta}_T^2} \, H_{3,7}^e
   + H_{3,8}^e
  \bigg)
 \bigg] \,. \label{e:gpd_gtmd_32}
\end{eqnarray}
The hermiticity constraint~(\ref{e:gtmd_hermiticity}) for the GTMDs, in combination 
with the relations~(\ref{e:gpd_gtmd_1})--(\ref{e:gpd_gtmd_32}), determines the 
symmetry behavior of the GPDs under the transformation $\xi \to - \xi$. 
One finds that the 10 GPDs $\tilde{E}_T$, $\tilde{H}_2$, $H'_2$, $E'_2$, 
$\tilde{E}'_2$, $H_{2T}$, $E_{2T}$, $\tilde{H}_{2T}$, $\tilde{E}'_{2T}$, 
$\tilde{E}_{3T}$ are odd functions in $\xi$, while all the other GPDs are even 
in $\xi$. 
This implies that the limit $\xi \to 0$ can be performed in 
eqs.~(\ref{e:gpd_gtmd_4}) and~(\ref{e:gpd_gtmd_28}) without encountering a 
singularity as the GPDs $\tilde{E}$ and $\tilde{E}_3$ are even functions in $\xi$. 
In addition, note that there appears no problem when performing the limit 
$\vec{\Delta}_T \to 0$ in eqs.~(\ref{e:gpd_gtmd_1})--(\ref{e:gpd_gtmd_32}) 
because of
\begin{eqnarray}
 \int d^2\vec{k}_T \, k_T^i \, 
 X(x,\xi,\vec{k}_T^2,\vec{k}_T \cdot \vec{\Delta}_T,\vec{\Delta}_T^2;\eta) 
 &\propto& \Delta_T^i \,, \\
 \int d^2\vec{k}_T \, (2 k_T^i k_T^j - \delta_T^{ij} \vec{k}_T^2) \, 
 X(x,\xi,\vec{k}_T^2,\vec{k}_T \cdot \vec{\Delta}_T,\vec{\Delta}_T^2;\eta) 
 &\propto& (2 \Delta_T^i \Delta_T^j - \delta_T^{ij} \vec{\Delta}_T^2) \,,
\end{eqnarray}
which holds for any GTMD $X$.

%
%
%
\subsection{Relations between GPDs and TMDs}
Having established the precise connection of the GPDs and TMDs with their 
respective {\it mother distributions} we are now in a position to search for 
possible model-independent relations between GPDs and TMDs. 
From~(\ref{e:tmd_gtmd_1}) and~(\ref{e:gpd_gtmd_1}) it is obvious that the 
GPD $H$ and the TMD $f_1$ can be related since both functions are projections 
of the GTMD $F_{1,1}^e$. 
With an analogous reasoning two additional relations can be obtained for 
twist-2, three for twist-3, and three for twist-4 leading altogether to  
\begin{eqnarray}
 H(x,0,0) 
 &=& \int d^2\vec{k}_T \, F_{1,1}^e(x,0,\vec{k}^2_T,0,0)
     = \int d^2 \vec{k}_T \, f_1(x,\vec{k}^2_T) \,, \label{e:trivial_1} \\
 \tilde{H}(x,0,0)
 &=& \int d^2\vec{k}_T \, G_{1,4}^e(x,0,\vec{k}^2_T,0,0)
     = \int d^2 \vec{k}_T \, g_{1L}(x,\vec{k}^2_T) \,, \label{e:trivial_2} \\
 H_T(x,0,0)
 &=& \int d^2\vec{k}_T \, \bigg[ H_{1,3}^e(x,0,\vec{k}^2_T,0,0)
     + \frac{\vec{k}_T^2}{2 M^2} \, H_{1,4}^e(x,0,\vec{k}^2_T,0,0) \bigg] \nonumber\\*
 &=& \int d^2 \vec{k}_T \, \bigg[ h_{1T}(x,\vec{k}^2_T)
     + \frac{\vec{k}_T^2}{2 M^2} \, h_{1T}^\bot(x,\vec{k}^2_T) \bigg]\,, \label{e:trivial_3}\\
 H_2(x,0,0)
 &=& \int d^2\vec{k}_T \, E_{2,1}^e(x,0,\vec{k}^2_T,0,0)
     = \int d^2 \vec{k}_T \, e(x,\vec{k}^2_T) \,, \\
 \tilde{H}'_2(x,0,0)
 &=& \int d^2\vec{k}_T \, H_{2,8}^e(x,0,\vec{k}^2_T,0,0)
     = \int d^2 \vec{k}_T \, h_L(x,\vec{k}^2_T) \,, \\
 H'_{2T}(x,0,0)
 &=& \int d^2\vec{k}_T \, \bigg[ G_{2,3}^e(x,0,\vec{k}^2_T,0,0)
     + \frac{\vec{k}_T^2}{2 M^2} \, G_{2,4}^e(x,0,\vec{k}^2_T,0,0) \bigg] \nonumber\\*
 &=& \int d^2 \vec{k}_T \, \bigg[ g'_T(x,\vec{k}^2_T)
     + \frac{\vec{k}_T^2}{2 M^2} \, g_{T}^\bot(x,\vec{k}^2_T) \bigg]\,, \\
 H_3(x,0,0)
 &=& \int d^2\vec{k}_T \, F_{3,1}^e(x,0,\vec{k}^2_T,0,0)
     = \int d^2 \vec{k}_T \, f_3(x,\vec{k}^2_T) \,, \\
 \tilde{H}_3(x,0,0)
 &=& \int d^2\vec{k}_T \, G_{3,4}^e(x,0,\vec{k}^2_T,0,0)
     = \int d^2 \vec{k}_T \, g_{3L}(x,\vec{k}^2_T) \,, \\
 H_{3T}(x,0,0)
 &=& \int d^2\vec{k}_T \, \bigg[ H_{3,3}^e(x,0,\vec{k}^2_T,0,0)
     + \frac{\vec{k}_T^2}{2 M^2} \, H_{3,4}^e(x,0,\vec{k}^2_T,0,0) \bigg] \nonumber\\*
 &=& \int d^2 \vec{k}_T \, \bigg[ h_{3T}(x,\vec{k}^2_T)
     + \frac{\vec{k}_T^2}{2 M^2} \, h_{3T}^\bot(x,\vec{k}^2_T) \bigg]\,.
\end{eqnarray}
These formulas can be considered as trivial model-independent relations between 
GPDs and TMDs (called relations of first type in ref.~\cite{Meissner:2007rx}).
Of course, the twist-2 relations~(\ref{e:trivial_1})--(\ref{e:trivial_3}) were already 
known before.

Here, we are mainly interested in nontrivial relations between GPDs and TMDs that have 
been suggested in the literature~\cite{Burkardt:2002ks,Burkardt:2003uw,Burkardt:2003je,Diehl:2005jf,Burkardt:2005hp,Lu:2006kt,Meissner:2007rx,Pasquini:2008ax}.
So far explicit relations have only been established in low-order calculations in the 
framework of simple spectator 
models~\cite{Burkardt:2003je,Burkardt:2005hp,Lu:2006kt,Meissner:2007rx}, and in one case 
in a light-cone constituent quark model~\cite{Pasquini:2008ax}. 
Our GTMD-analysis can now shed light on the question if model-independent 
nontrivial relations exist.
%
\FIGURE[t]{%
 \includegraphics{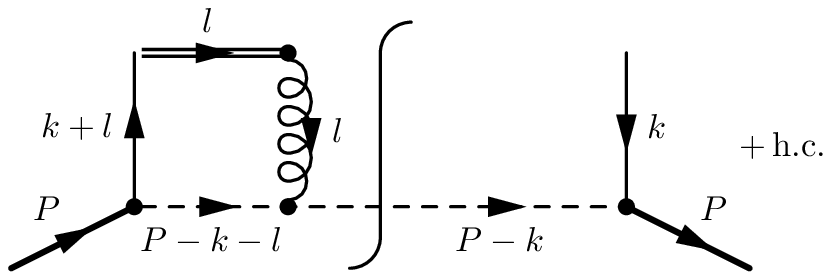}
 \caption{Lowest nontrivial order diagram for T-odd TMDs in the scalar diquark 
spectator model. The Hermitian conjugate diagram (h.c.) is not shown. The eikonal 
propagator arising from the Wilson line in the operator definition of TMDs is 
indicated by a double line.}
 \label{f:todd}}
%

A complete classification of the nontrivial relations between GPDs and TMDs in leading 
twist has been performed in~\cite{Meissner:2007rx}, where explicit formulae have been
obtained in the same diquark spectator model as discussed in 
appendix~\ref{c:app_gtmd_model}. 
In that work two distinct types of nontrivial relations between quark distributions
have been considered --- one connecting certain GPDs with the T-odd\footnote{Note that 
in order to generate T-odd TMDs one has to take into account rescattering effects 
between the active parton and the spectator system. Therefore, in the diquark spectator
model the lowest order contribution to T-odd TMDs comes from the diagram shown in 
figure~\ref{f:todd}.} Sivers function $f_{1T}^\bot$~\cite{Sivers:1989cc,Sivers:1990fh}
and the Boer-Mulders function $h_1^\bot$~\cite{Boer:1997nt} (called relations 
of second type in ref.~\cite{Meissner:2007rx}),
\begin{eqnarray}
 E(x,0,-\vec{\Delta}_T^2) &\leftrightarrow& -f_{1T}^\bot(x,\vec{k}_T^2;\eta) \,, \label{e:model_rel_1} \\
 E_T(x,0,-\vec{\Delta}_T^2) + 2\tilde{H}_T(x,0,-\vec{\Delta}_T^2) &\leftrightarrow& -h_1^\bot(x,\vec{k}_T^2;\eta)
 \,, \label{e:model_rel_2}
\end{eqnarray}
and one connecting a GPD and the T-even pretzelosity TMD $h_{1T}^\bot$ 
(called relation of third type in ref.~\cite{Meissner:2007rx}),
\begin{eqnarray}
 \tilde{H}_T(x,0,-\vec{\Delta}_T^2) &\leftrightarrow& \tfrac{1}{2} \, h_{1T}^\bot(x,\vec{k}_T^2)
 \,. \label{e:model_rel_3}
\end{eqnarray}
As we discuss in the following our GTMD-analysis, however, does not support a 
model-independent status of any such relations.

For the relations of second type in eqs.~(\ref{e:model_rel_1}) 
and~(\ref{e:model_rel_2}) this is obvious because, according to 
eqs.~(\ref{e:tmd_gtmd_2}), (\ref{e:tmd_gtmd_5}), (\ref{e:gpd_gtmd_2}),
(\ref{e:gpd_gtmd_6}), and~(\ref{e:gpd_gtmd_7}) the involved GPDs and TMDs have 
different, independent {\it mother distributions}.
In particular, the GPDs are connected to T-even parts of GTMDs while the TMDs are 
connected to T-odd parts of GTMDs. 
Unless, for some reason, the GTMDs are subject to further constraints one has to 
conclude that there cannot exist a model-independent relation between the GPDs and 
TMDs given in eqs.~(\ref{e:model_rel_1}) and~(\ref{e:model_rel_2}). 
This conclusion is in accordance with the observation made in~\cite{Meissner:2007rx} 
that nontrivial relations of second type are likely to even break down in spectator 
models once higher order contributions are taken into account. 
Therefore, one has to attribute the relations to the simplicity of the used model. 
Nevertheless, it may well be that numerically the model-dependent nontrivial relations 
work reasonably well when comparing to experimental data. 
In fact such a case is already known for distributions of the nucleon, namely the 
relation between the Sivers function and the GPD 
$E$~\cite{Burkardt:2002ks,Burkardt:2003uw,Burkardt:2003je,Meissner:2007rx}.

For the relation of third type in eq.~(\ref{e:model_rel_3}) the GPD as well as
the TMD are, according to eqs.~(\ref{e:tmd_gtmd_8}) and~(\ref{e:gpd_gtmd_7}),
related to T-even parts of GTMDs. 
But the linear combinations of GTMDs differ in both cases such that no 
model-independent nontrivial relation of the type~(\ref{e:model_rel_3}) 
can exist.
In the context of the diquark spectator model the explicit relation
\begin{equation} \label{e:rel_expl}
 \frac{3}{(1-x)^2} \, \tilde{H}_T(x,0,0)
 = \int d^2 \vec{k}_T \, h_{1T}^\bot(x,\vec{k}^2_T)\,,
\end{equation} 
was established~\cite{Meissner:2007rx}.
One may wonder if, in general, the specific kinematical point 
$\vec{\Delta}_T^2 = \xi = 0$ and the $k_T$-integration used in~(\ref{e:rel_expl})
might spoil the above argument about different linear combinations of GTMDs.
However, by taking all known symmetry properties of the GTMDs into account one 
is still left with such different linear combinations.
Even in the simple diquark spectator model this is the case, and the 
relation~(\ref{e:rel_expl}) also just holds due to the simplicity of the model.

In order to illustrate this point we calculate the involved GPD $\tilde{H}_T$ and 
TMD $h_{1T}^\bot$ in the scalar diquark model and try to preserve their respective 
GTMD structure as far as possible. 
By inserting the model results for the GTMDs in appendix~\ref{c:app_gtmd_model} 
into eq.~(\ref{e:gpd_gtmd_7}) one finds for the GPD $\tilde{H}_T$ in the case 
$\xi=0$
\begin{align}
 &\tilde{H}_T(x,0,-\vec{\Delta}_T^2) \nonumber\\*
 &\quad= \int d^2\vec{k}_T \, \tilde{C} \, \bigg[
  \tilde{H}_{1,2}^e(x)
  - 2 \bigg(
   \frac{2 (\vec{k}_T \cdot \vec{\Delta}_T)^2 - \vec{k}_T^2 \vec{\Delta}_T^2}{(\vec{\Delta}_T^2)^2} \,
    \tilde{H}_{1,4}^e(x)
   + \tilde{H}_{1,6}^e(x)
  \bigg)
 \bigg] \,.
\end{align}
Here we have extracted all dependence on the vectors $\vec{k}_T$ and $\vec{\Delta}_T$ from the GTMDs and put it either into their coefficients or into the overall factor
\begin{equation}
 \tilde{C}
 = \frac{g^2 \, (1-x)}{2(2\pi)^3} \, \frac{1}{
   [(\vec{k}_T + \tfrac{1}{2}(1-x) \, \vec{\Delta}_T)^2 + \tilde{M}^2(x)] \,
   [(\vec{k}_T - \tfrac{1}{2}(1-x) \, \vec{\Delta}_T)^2 + \tilde{M}^2(x)]} \,,
\end{equation}
with
\begin{equation}
 \tilde{M}^2(x) = x \, m_s^2 + (1-x) \, m_q^2 - x (1-x) \, M^2 \,.
\end{equation}
Therefore, the remnants of the GTMDs
\begin{eqnarray}
 \tilde{H}_{1,2}^e(x) & = & (1-x) \, (m_q + x M) M\,, \\
 \tilde{H}_{1,4}^e(x) & = & -2 M^2\,, \\
 \tilde{H}_{1,6}^e(x) & = & \tfrac{1}{2}(1-x) \, (m_q + M) M
\end{eqnarray}
only depend on the momentum fraction $x$. 
This allows one to perform the $\vec{k}_T$ integration, which yields
\begin{align}
 &\tilde{H}_T(x,0,-\vec{\Delta}_T^2) \nonumber\\*
 &\quad= \frac{g^2 \, (1-x)}{8(2\pi)^2} \, \int_0^1 d\alpha \, \frac{
  2 \tilde{H}_{1,2}^e(x)
  - (1 - 2\alpha)^2 \, (1-x)^2 \, \tilde{H}_{1,4}^e(x)
  - 4 \tilde{H}_{1,6}^e(x)
 }{\alpha(1-\alpha) \, (1-x)^2 \, \vec{\Delta}_T^2 + \tilde{M}^2(x)} \,.
\end{align}
In the forward limit this leads to
\begin{equation}
 \tilde{H}_T(x,0,0) = \frac{g^2 \, (1-x)}{8(2\pi)^2} \, \frac{
  2 \tilde{H}_{1,2}^e(x)
  - \tfrac{1}{3} (1-x)^2 \, \tilde{H}_{1,4}^e(x)
  - 4 \tilde{H}_{1,6}^e(x)
 }{\tilde{M}^2(x)} \,.
\end{equation}
On the other hand, one finds for the zeroth moment of the TMD $h_{1T}^\bot$ by 
inserting the model results for the GTMDs in appendix~\ref{c:app_gtmd_model} 
into eq.~(\ref{e:tmd_gtmd_8})
\begin{equation}
 \int d^2 \vec{k}_T \, h_{1T}^\bot(x,\vec{k}^2_T)
 = \frac{g^2 \, (1-x)}{4(2\pi)^2} \, \frac{\tilde{H}_{1,4}^e(x)}{\tilde{M}^2(x)} \,.
\end{equation}
This shows explicitly that the GPD $\tilde{H}_T$ and the TMD $h_{1T}^\bot$ are 
connected to different remnants of GTMDs even in the scalar diquark model. 

However, due to the simplicity of the scalar diquark model the remnants of the 
GTMDs are related according to
\begin{equation}
 2\tilde{H}_{1,2}^e(x) - 4\tilde{H}_{1,6}^e(x) = -2 (1-x)^2 M^2 = (1-x)^2 \, \tilde{H}_{1,4}^e(x) \,.
 \label{e:gtmd_rel}
\end{equation}
This immediately implies the relation
\begin{equation}
 \frac{3}{(1-x)^2} \, \tilde{H}_T(x,0,0)
 = \frac{g^2 \, (1-x)}{4(2\pi)^2} \, \frac{\tilde{H}_{1,4}^e(x)}{\tilde{M}^2(x)}
 = \int d^2 \vec{k}_T \, h_{1T}^\bot(x,\vec{k}^2_T)\,,
\end{equation}
which we already quoted above in~(\ref{e:rel_expl}). 
It should be stressed once again, that this relation only holds due to the 
simplicity of the scalar diquark model. 
In general, no dependence like eq.~(\ref{e:gtmd_rel}) will exist between the 
different, independent GTMDs.
We note that a relation like~(\ref{e:rel_expl}) was also obtained in a specific 
light-cone quark model~\cite{Pasquini:2008ax}, but in that model a factor different 
from 3 on the {\it l.h.s.}~of~(\ref{e:rel_expl}) shows up\footnote{Actually,
in ref.~\cite{Pasquini:2008ax} the factor 3 appeared, but later on an error 
in the calculation was found~\cite{Pasquini:2009}.}.
The fact that a formula corresponding to~(\ref{e:rel_expl}) emerges in the 
framework of another model does not contradict our general argument that in 
full QCD a relation of the type~(\ref{e:model_rel_3}) cannot hold.

By extending our GTMD analysis we find that also for twist-3 and twist-4 no
model-independent nontrivial relations between GPDs and TMDs exist.
On the other hand such relations may well emerge in the framework of simple
models.

%
%
%
\section{Conclusions}\label{c:sec5}
In summary, we have derived the structure of the fully unintegrated, off-diagonal 
quark-quark correlator for a spin-1/2 hadron, and thus extended our previous
study of the spin-0 case~\cite{Meissner:2008ay}. 
This object, which contains the most general information on the two-parton 
structure of a hadron, has been parameterized in terms of so-called generalized
parton correlation functions (GPCFs).
The major challenge in this derivation was to eliminate all redundant terms
without missing any relevant term at the same time.
Integrating the GPCFs upon a light-cone component of the quark momentum one ends
up with entities which we called generalized transverse momentum dependent parton 
distributions (GTMDs).
In general, GTMDs can be of direct relevance for the phenomenology of various hard 
(diffractive) processes 
(see, e.g., refs.~\cite{Martin:1999wb,Khoze:2000cy,Goloskokov:2007nt,Albrow:2008pn}).
Our analysis shows that both the GPCFs and the GTMDs in general are complex-valued 
functions.
This is different from the (simpler) forward parton distributions, GPDs, and TMDs 
all of which are real.

Suitable projections of GTMDs lead to GPDs on the one hand and TMDs on the other.
Therefore, GTMDs can be considered as {\it mother distributions} of GPDs and 
TMDs~\cite{Ji:2003ak,Belitsky:2003nz,Belitsky:2005qn}. 
To study these two limiting cases of GTMDs was the main motivation of the present 
work.
One outcome was the first complete classification of GPDs for a spin-1/2 hadron
beyond leading twist.
Most importantly, we were able to determine which of the GPDs and TMDs have the 
same {\it mother distributions} allowing us to explore whether model-independent 
relations between GPDs and TMDs can be established.
One ends up with nine such model-independent relations.
Actually, these cases can be considered as trivial ones because the respective 
GPDs and TMDs also have a relation to the same forward parton distributions 
(see also ref.~\cite{Meissner:2007rx}).
Our main interest was to investigate nontrivial relations between GPDs and TMDs 
which have been obtained in models and extensively discussed in the recent 
literature~\cite{Burkardt:2002ks,Burkardt:2003uw,Burkardt:2003je,Diehl:2005jf,Burkardt:2005hp,Lu:2006kt,Meissner:2007rx,Pasquini:2008ax}.
We have restricted this study to leading twist where three nontrivial relations 
have been found (see~\cite{Meissner:2007rx} for an overview)
--- two involving the T-odd Sivers TMD $f_{1T}^\bot$ and the Boer-Mulders TMD 
$h_1^\bot$, and one in which the T-even pretzelosity TMD $h_{1T}^\bot$ shows
up.
It turns out that none of these relations can be promoted to a model-independent 
status as the respective functions are related to different (linear combinations of) 
GTMDs.
For the relations containing T-odd TMDs this finding agrees with 
ref.~\cite{Meissner:2007rx} where it has been argued that these nontrivial 
relations between GPDs and TMDs are likely to break down even in spectator 
models if the parton distributions are evaluated to higher order in perturbation 
theory.
Moreover, our model-independent study for the Boer-Mulders function of a spin-0 
hadron came to the same conclusion~\cite{Meissner:2008ay}.
We emphasize that our finding does not tell anything about the numerical 
violation of (model-dependent) nontrivial relations between GPDs and TMDs. 
On the other hand, such relations have hardly any predictive power and only 
after all the involved distributions have been measured one can really judge
about their quality.

%
%
%
\acknowledgments 
The work has partially been supported by the Verbundforschung ``Hadronen und Kerne'' 
of the BMBF and by the Deutsche Forschungsgemeinschaft (DFG).
\\[0.3cm]
\noindent
\textbf{Notice:} Authored by Jefferson Science Associates, LLC under U.S. DOE 
Contract No. DE-AC05-06OR23177. 
The U.S. Government retains a non-exclusive, paid-up, irrevocable, world-wide license 
to publish or reproduce this manuscript for U.S. Government purposes. 

%
%
%
\appendix
\section{Parameterization of Dirac bilinears}\label{c:app_par}
In this appendix we derive the most general parameterization of the scalar, 
pseudoscalar, vector, axial vector, and tensor Dirac bilinear introduced in the 
eqs.~(\ref{e:sdb})--(\ref{e:tdb}) respecting the corresponding constraints from parity
\begin{eqnarray}
 \Gamma_\text{S}(P, k, \Delta, N; \eta) &=&
 + \gamma_0 \, \Gamma_\text{S}(\bar{P}, \bar{k}, \bar{\Delta}, \bar{N}; \eta) \, \gamma_0
 \label{e:ssp} \,,\\
 \Gamma_\text{P}(P, k, \Delta, N; \eta) &=&
 - \gamma_0 \, \Gamma_\text{P}(\bar{P}, \bar{k}, \bar{\Delta}, \bar{N}; \eta) \, \gamma_0
 \label{e:psp} \,,\\
 \Gamma^\mu_\text{V}(P, k, \Delta, N; \eta) &=&
 + \gamma_0 \, \Gamma^{\bar{\mu}}_\text{V}(\bar{P}, \bar{k}, \bar{\Delta}, \bar{N}; \eta) \, \gamma_0
 \label{e:vsp} \,,\\
 \Gamma^\mu_\text{A}(P, k, \Delta, N; \eta) &=&
 - \gamma_0 \, \Gamma^{\bar{\mu}}_\text{A}(\bar{P}, \bar{k}, \bar{\Delta}, \bar{N}; \eta) \, \gamma_0
 \label{e:asp} \,,\\
 \Gamma^{\mu\nu}_\text{T}(P, k, \Delta, N; \eta) &=&
 + \gamma_0 \, \Gamma^{\bar{\mu}\bar{\nu}}_\text{T}(\bar{P}, \bar{k}, \bar{\Delta}, \bar{N}; \eta) \, \gamma_0
 \label{e:tsp} \,.
\end{eqnarray}
For this purpose we generalize the method described in~\cite{Diehl:2001pm} for GPDs 
in an appropriate way to the case of GPCFs.
In order to eliminate independent terms in the parameterization of the Dirac bilinears 
we use the Gordon identities
\begin{eqnarray}
 \bar{u}(p', \lambda') \, \gamma^\mu \, u(p, \lambda)
 &=& \bar{u}(p', \lambda') \, \bigg[ \frac{P^\mu}{M}
     + \frac{i \sigma^{\mu\Delta}}{2M} \bigg] \, u(p, \lambda) \label{e:gi1} \,,\\
 0
 &=& \bar{u}(p', \lambda') \, \bigg[ \frac{\Delta^\mu}{2M}
     + \frac{i \sigma^{\mu P}}{M} \bigg] \, u(p, \lambda) \label{e:gi2} \,,\\
 \bar{u}(p', \lambda') \, \gamma^\mu \gamma_5 \, u(p, \lambda)
 &=& \bar{u}(p', \lambda') \, \bigg[ \frac{\Delta^\mu \gamma_5}{2M}
     + \frac{i \sigma^{\mu P} \gamma_5}{M} \bigg] \, u(p, \lambda) \label{e:gi3} \,,\\
 0
 &=& \bar{u}(p', \lambda') \, \bigg[ \frac{P^\mu \gamma_5}{M}
     + \frac{i \sigma^{\mu\Delta} \gamma_5}{2M} \bigg] \, u(p, \lambda) \label{e:gi4} \,.
\end{eqnarray}
In addition, we also use the $\varepsilon$-identity
\begin{equation}
 g^{\alpha\beta} \varepsilon^{\mu\nu\rho\sigma}
 = g^{\mu\beta} \varepsilon^{\alpha\nu\rho\sigma}
 + g^{\nu\beta} \varepsilon^{\mu\alpha\rho\sigma}
 + g^{\rho\beta} \varepsilon^{\mu\nu\alpha\sigma}
 + g^{\sigma\beta} \varepsilon^{\mu\nu\rho\alpha}
 \label{e:ei}
\end{equation}
as well as the $\sigma$-identity
\begin{equation}
 i\sigma^{\mu\nu} \gamma_5 = - \tfrac{1}{2} \varepsilon^{\mu\nu\rho\sigma} \sigma_{\rho\sigma} \label{e:si} \,.
\end{equation}

%
%
%
\subsection{Parameterization of the scalar Dirac bilinear}
A complete parameterization of the scalar Dirac bilinear in eq.~(\ref{e:sdb}) can be obtained by treating all possible Dirac currents one after the other:
\begin{enumerate}
\item {\it vector current} [$\bar{u}(p', \lambda') \, \gamma^\mu \, u(p, \lambda)$]: Using the Gordon identity in eq.~(\ref{e:gi1}) all vector currents can be replaced by scalar and tensor currents.
\item {\it axial vector current} [$\bar{u}(p', \lambda') \, \gamma^\mu \gamma_5 \, u(p, \lambda)$]: Using the Gordon identity in eq.~(\ref{e:gi3}) all axial vector currents can be replaced by pseudoscalar and pseudotensor currents.
\item {\it pseudoscalar current} [$\bar{u}(p', \lambda') \, \gamma_5 \, u(p, \lambda)$]: Contracting the Gordon identity in eq.~(\ref{e:gi4}) with the light-cone vector $N$ yields
\begin{equation}
 \bar{u}(p', \lambda') \, \gamma_5 \, u(p, \lambda)
 = \bar{u}(p', \lambda') \, \frac{i \sigma^{\Delta N} \gamma_5}{2 M^2} \, u(p, \lambda) \label{e:gi4a} \,,
\end{equation}
so that all pseudoscalar currents can be replaced by pseudotensor currents.
\item {\it pseudotensor current} [$\bar{u}(p', \lambda') \, i\sigma^{\mu\nu} \gamma_5 \, u(p, \lambda)$]: Using the $\sigma$-identity in eq.~(\ref{e:si}) all pseudotensor currents can be replaced by tensor currents.
\item {\it tensor current} [$\bar{u}(p', \lambda') \, i\sigma^{\mu\nu} \, u(p, \lambda)$]: All possible tensor currents are of the form
\begin{equation}
 \bar{u}(p', \lambda') \, i\sigma^{ab} \, u(p, \lambda)
\end{equation}
with $a$ and $b$ being any of the vectors $P$, $k$, $\Delta$, and $N$. Using the Gordon identity in eq.~(\ref{e:gi2}) all tensor currents containing the vector $P$ contracted with $\sigma$ can be replaced by scalar currents. Therefore, one is left with three tensor currents
\begin{equation}
 \bar{u}(p', \lambda') \, i\sigma^{k\Delta} \, u(p, \lambda) \,, \quad
 \bar{u}(p', \lambda') \, i\sigma^{kN} \, u(p, \lambda) \,, \quad
 \bar{u}(p', \lambda') \, i\sigma^{\Delta N} \, u(p, \lambda) \label{e:sdb1} \,.
\end{equation}
\item {\it scalar current} [$\bar{u}(p', \lambda') \, u(p, \lambda)$]: There is only one possible scalar current
\begin{equation}
 \bar{u}(p', \lambda') \, u(p, \lambda) \label{e:sdb2} \,,
\end{equation}
which can not be replaced.
\end{enumerate}
To summarize, the scalar Dirac bilinear in eq.~(\ref{e:sdb}) is completely parameterized by the four currents in eqs.~(\ref{e:sdb1}) and~(\ref{e:sdb2}), i.~e.,
\begin{align}
 &\bar{u}(p', \lambda') \, \Gamma_\text{S}(P, k, \Delta, N; \eta) \, u(p, \lambda) \nonumber\\
 &\quad = \bar{u}(p', \lambda') \, \bigg[
  A^E_{1}
  + \frac{i\sigma^{k\Delta}}{M^2} \, A^E_{2}
  + \frac{i\sigma^{kN}}{M^2} \, A^E_{3}
  + \frac{i\sigma^{\Delta N}}{M^2} \, A^E_{4} \bigg] \, u(p, \lambda) \label{e:psdb} \,,
\end{align}
where the GPCFs $A^E_n$ are scalar functions of $P$, $k$, $\Delta$, $N$, and $\eta$.

%
%
%
\subsection{Parameterization of the pseudoscalar Dirac bilinear}
A complete parameterization of the pseudoscalar Dirac bilinear in eq.~(\ref{e:pdb}) can be obtained by treating all possible Dirac currents one after the other:
\begin{enumerate}
\item {\it vector current} [$\bar{u}(p', \lambda') \, \gamma^\mu \, u(p, \lambda)$]: Using the Gordon identity in eq.~(\ref{e:gi1}) all vector currents can be replaced by scalar and tensor currents.
\item {\it axial vector current} [$\bar{u}(p', \lambda') \, \gamma^\mu \gamma_5 \, u(p, \lambda)$]: Using the Gordon identity in eq.~(\ref{e:gi3}) all axial vector currents can be replaced by pseudoscalar and pseudotensor currents.
\item {\it pseudoscalar current} [$\bar{u}(p', \lambda') \, \gamma_5 \, u(p, \lambda)$]: Using eq.~(\ref{e:gi4a}) all pseudoscalar currents can be replaced by pseudotensor currents.
\item {\it tensor current} [$\bar{u}(p', \lambda') \, i\sigma^{\mu\nu} \, u(p, \lambda)$]: Using the $\sigma$-identity in eq.~(\ref{e:si}) all tensor currents can be replaced by pseudotensor currents.
\item {\it pseudotensor current} [$\bar{u}(p', \lambda') \, i\sigma^{\mu\nu} \gamma_5 \, u(p, \lambda)$]: All possible pseudotensor currents are of the form
\begin{equation}
 \bar{u}(p', \lambda') \, i\sigma^{ab} \gamma_5 \, u(p, \lambda)
\end{equation}
with $a$ and $b$ being any of the vectors $P$, $k$, $\Delta$, and $N$. Now, multiplying the Gordon identity in eq.~(\ref{e:gi2}) with an $\varepsilon$-tensor and using the $\varepsilon$-identity in eq.~(\ref{e:ei}) as well as the $\sigma$-identity in eq.~(\ref{e:si}) yields
\begin{equation}
 0 = \bar{u}(p', \lambda') \, \bigg[ \frac{P^\mu \, i\sigma^{\nu\rho} \gamma_5}{M}
     + \frac{P^\nu \, i\sigma^{\rho\mu} \gamma_5}{M}
     + \frac{P^\rho \, i\sigma^{\mu\nu} \gamma_5}{M}
     - \frac{i\varepsilon^{\mu\nu\rho\Delta}}{2M} \bigg] \, u(p, \lambda) \label{e:gi2a} \,.
\end{equation}
Contracting this equation in turn with $P_\mu$, $k_\nu$, $N_\rho$ and $P_\mu$, $\Delta_\nu$, $N_\rho$ and $k_\mu$, $\Delta_\nu$, $N_\rho$ allows to replace some pseudotensor currents by scalar and pseudotensor currents
\begin{align}
 &\bar{u}(p', \lambda') \, i\sigma^{Pk} \gamma_5 \, u(p, \lambda) \nonumber\\*
 &\quad = \bar{u}(p', \lambda') \, \bigg[
   \frac{P \cdot k \, i\sigma^{PN} \gamma_5}{M^2}
   - \frac{P^2 \, i\sigma^{kN} \gamma_5}{M^2}
   - \frac{i\varepsilon^{Pk\Delta N}}{2 M^2}
  \bigg] \, u(p, \lambda) \label{e:gi2b} \,, \\
 &\bar{u}(p', \lambda') \, i\sigma^{P\Delta} \gamma_5 \, u(p, \lambda) \nonumber\\*
 &\quad = \bar{u}(p', \lambda') \, \bigg[
   - \frac{P^2 \, i\sigma^{\Delta N} \gamma_5}{M^2}
  \bigg] \, u(p, \lambda) \label{e:gi2c} \,, \\
 &\bar{u}(p', \lambda') \, i\sigma^{k\Delta} \gamma_5 \, u(p, \lambda) \nonumber\\*
 &\quad = \bar{u}(p', \lambda') \, \bigg[
   - \frac{P \cdot k \, i\sigma^{\Delta N} \gamma_5}{M^2}
  \bigg] \, u(p, \lambda) \label{e:gi2d} \,,
\end{align}
so that one is left with three pseudotensor currents
\begin{equation}
 \bar{u}(p', \lambda') \, i\sigma^{PN} \gamma_5 \, u(p, \lambda) \,, \quad
 \bar{u}(p', \lambda') \, i\sigma^{kN} \gamma_5 \, u(p, \lambda) \,, \quad
 \bar{u}(p', \lambda') \, i\sigma^{\Delta N} \gamma_5 \, u(p, \lambda) \label{e:pdb1} \,.\!
\end{equation}
\item {\it scalar current} [$\bar{u}(p', \lambda') \, u(p, \lambda)$]: There is only one possible scalar current
\begin{equation}
 \bar{u}(p', \lambda') \, i\varepsilon^{Pk\Delta N} \, u(p, \lambda) \label{e:pdb2} \,,
\end{equation}
which can not be replaced.
\end{enumerate}
To summarize, the pseudoscalar Dirac bilinear in eq.~(\ref{e:pdb}) is completely parameterized by the four currents in eqs.~(\ref{e:pdb1}) and~(\ref{e:pdb2}), i.~e.,
\begin{align}
 &\bar{u}(p', \lambda') \, \Gamma_\text{P}(P, k, \Delta, N; \eta) \, u(p, \lambda) \nonumber\\
 &\quad = \bar{u}(p', \lambda') \, \bigg[
  \frac{i\varepsilon^{Pk\Delta N}}{M^4} \, A^E_{5}
  + \frac{i\sigma^{PN} \gamma_5}{M^2} \, A^E_{6}
  + \frac{i\sigma^{kN} \gamma_5}{M^2} \, A^E_{7}
  + \frac{i\sigma^{\Delta N} \gamma_5}{M^2} \, A^E_{8} \bigg] \, u(p, \lambda) \label{e:ppdb} \,,
\end{align}
where the GPCFs $A^E_n$ are scalar functions of $P$, $k$, $\Delta$, $N$, and $\eta$.

%
%
%
\subsection{Parameterization of the vector Dirac bilinear}
A complete parameterization of the vector Dirac bilinear in eq.~(\ref{e:vdb}) can be obtained by treating all possible Dirac currents one after the other:
\begin{enumerate}
\item {\it vector current} [$\bar{u}(p', \lambda') \, \gamma^\mu \, u(p, \lambda)$]: Using the Gordon identity in eq.~(\ref{e:gi1}) all vector currents can be replaced by scalar and tensor currents.
\item {\it axial vector current} [$\bar{u}(p', \lambda') \, \gamma^\mu \gamma_5 \, u(p, \lambda)$]: Using the Gordon identity in eq.~(\ref{e:gi3}) all axial vector currents can be replaced by pseudoscalar and pseudotensor currents.
\item {\it pseudoscalar current} [$\bar{u}(p', \lambda') \, \gamma_5 \, u(p, \lambda)$]: Using eq.~(\ref{e:gi4a}) all pseudoscalar currents can be replaced by pseudotensor currents.
\item {\it pseudotensor current} [$\bar{u}(p', \lambda') \, i\sigma^{\mu\nu} \gamma_5 \, u(p, \lambda)$]: Using the $\sigma$-identity in eq.~(\ref{e:si}) all pseudotensor currents can be replaced by tensor currents.
\item {\it tensor current} [$\bar{u}(p', \lambda') \, i\sigma^{\mu\nu} \, u(p, \lambda)$]: All possible tensor currents are of the form
\begin{equation}
 \bar{u}(p', \lambda') \, i\sigma^{\mu a} \, u(p, \lambda) \,,\quad
 \bar{u}(p', \lambda') \, a^\mu \, i\sigma^{bc} \, u(p, \lambda)
\end{equation}
with $a$, $b$, and $c$ being any of the vectors $P$, $k$, $\Delta$, and $N$. Using the Gordon identity in eq.~(\ref{e:gi2}) all tensor currents containing the vector $P$ contracted with $\sigma$ can be replaced by scalar currents. Therefore, one is left with 15 tensor currents
\begin{align}
 &\bar{u}(p', \lambda') \, i\sigma^{\mu k} \, u(p, \lambda) \,, &
 &\bar{u}(p', \lambda') \, i\sigma^{\mu\Delta} \, u(p, \lambda) \,, &
 &\bar{u}(p', \lambda') \, i\sigma^{\mu N} \, u(p, \lambda) \,, \nonumber\\*
 &\bar{u}(p', \lambda') \, a^\mu \, i\sigma^{k\Delta} \, u(p, \lambda) \,, &
 &\bar{u}(p', \lambda') \, a^\mu \, i\sigma^{kN} \, u(p, \lambda) \,, &
 &\bar{u}(p', \lambda') \, a^\mu \, i\sigma^{\Delta N} \, u(p, \lambda) \!\! \label{e:vdb1}
\end{align}
with $a$ being any of the vectors $P$, $k$, $\Delta$, and $N$.
\item {\it scalar current} [$\bar{u}(p', \lambda') \, u(p, \lambda)$]: There are four possible scalar currents
\begin{equation}
 \bar{u}(p', \lambda') \, a^\mu \, u(p, \lambda) \label{e:vdb2}
\end{equation}
with $a$ being any of the vectors $P$, $k$, $\Delta$, and $N$, which can not be replaced.
\end{enumerate}
So far, we were able to reduce the number of currents needed to parameterize the vector Dirac bilinear in eq.~(\ref{e:vdb}) to the 19 currents in eqs.~(\ref{e:vdb1}) and~(\ref{e:vdb2}). However, it is possible to reduce this number even further by using that in Minkowski space
\begin{equation}
 \det\left(\begin{array}{ccccc}
  g^{\alpha\mu} & \ g^{\beta\mu} \ & g^{\gamma\mu} & \ g^{\delta\mu} \ & g^{\varepsilon\mu} \\
  g^{\alpha\nu} & \ g^{\beta\nu} \ & g^{\gamma\nu} & \ g^{\delta\nu} \ & g^{\varepsilon\nu} \\
  g^{\alpha\rho} & \ g^{\beta\rho} \ & g^{\gamma\rho} & \ g^{\delta\rho} \ & g^{\varepsilon\rho} \\
  g^{\alpha\sigma} & \ g^{\beta\sigma} \ & g^{\gamma\sigma} & \ g^{\delta\sigma} & g^{\varepsilon\sigma} \\
  g^{\alpha\tau} & \ g^{\beta\tau} \ & g^{\gamma\tau} & \ g^{\delta\tau} \ & g^{\varepsilon\tau} \\
 \end{array}\right)=0 \label{e:di} \,.
\end{equation}
Contracting this determinant with $P_\alpha$, $k_\beta$, $\Delta_\gamma$, $N_\delta$, $P_\nu$, $k_\rho$, $\Delta_\sigma$, and $N_\tau$ and then in turn with $i{\sigma_\varepsilon}^k$, $i{\sigma_\varepsilon}^\Delta$, and $i{\sigma_\varepsilon}^N$ yields
\begin{eqnarray}
 \bar{u}(p', \lambda') \, \det\left(\begin{array}{ccccc}
  P^\mu     & \ k^\mu \          & \Delta^\mu     & \ N^\mu \          & i\sigma^{\mu k} \\
  P^2       & \ P \cdot k \      & 0              & \ P \cdot N \      & \frac{1}{2} k \cdot \Delta \\
  P \cdot k & \ k^2 \            & k \cdot \Delta & \ k \cdot N \      & 0 \\
  0         & \ k \cdot \Delta \ & \Delta^2       & \ \Delta \cdot N \ & -i\sigma^{k\Delta} \\
  P \cdot N & \ k \cdot N \      & \Delta \cdot N & \ 0 \              & -i\sigma^{kN}
 \end{array}\right) \, u(p, \lambda) &=& 0 \label{e:vdb3a} \,, \\
 \bar{u}(p', \lambda') \, \det\left(\begin{array}{ccccc}
  P^\mu     & \ k^\mu \          & \Delta^\mu     & \ N^\mu \          & i\sigma^{\mu\Delta} \\
  P^2       & \ P \cdot k \      & 0              & \ P \cdot N \      & \frac{1}{2} \Delta^2 \\
  P \cdot k & \ k^2 \            & k \cdot \Delta & \ k \cdot N \      & i\sigma ^{k\Delta} \\
  0         & \ k \cdot \Delta \ & \Delta^2       & \ \Delta \cdot N \ & 0 \\
  P \cdot N & \ k \cdot N \      & \Delta \cdot N & \ 0 \              & -i\sigma^{\Delta N}
 \end{array}\right) \, u(p, \lambda) &=& 0 \label{e:vdb3b} \,, \\
 \bar{u}(p', \lambda') \, \det\left(\begin{array}{ccccc}
  P^\mu     & \ k^\mu \          & \Delta^\mu     & \ N^\mu \          & i\sigma^{\mu N} \\
  P^2       & \ P \cdot k \      & 0              & \ P \cdot N \      & \frac{1}{2} \Delta \cdot N \\
  P \cdot k & \ k^2 \            & k \cdot \Delta & \ k \cdot N \      & i\sigma^{kN} \\
  0         & \ k \cdot \Delta \ & \Delta^2       & \ \Delta \cdot N \ & i\sigma^{\Delta N} \\
  P \cdot N & \ k \cdot N \      & \Delta \cdot N & \ 0 \              & 0
 \end{array}\right) \, u(p, \lambda) &=& 0 \label{e:vdb3c} \,.
\end{eqnarray}
For $P\not=0$, $k\not=0$, $\Delta\not=0$, and $N\not=0$ these three equations are non-trivial and allow one to eliminate three of the 19 currents we have left. As $P$ and $N$ are always different from zero, it is most convenient to eliminate the tensor currents
\begin{equation}
 \Delta^\mu \, i\sigma^{k\Delta} \,,\quad
 \Delta^\mu \, i\sigma^{kN} \,,\quad
 k^\mu \, i\sigma^{\Delta N}
\end{equation}
because either $k$ and $\Delta$ are different form zero and we are able to eliminate them using the constraints in eqs.~(\ref{e:vdb3a})--(\ref{e:vdb3c}) or at least one of the vectors is zero and the whole current vanishes. Therefore, one possible parameterization of the vector Dirac bilinear in eq.~(\ref{e:vdb}), is given by
\begin{align}
 &\bar{u}(p', \lambda') \, \Gamma^\mu_\text{V}(P, k, \Delta, N; \eta) \, u(p, \lambda) \nonumber\\
 &\quad = \bar{u}(p', \lambda') \, \bigg[
  \frac{P^\mu}{M} \, A^F_{1}
  + \frac{k^\mu}{M} \, A^F_{2}
  + \frac{\Delta^\mu}{M} \, A^F_{3}
  + \frac{N^\mu}{M} \, A^F_{4}
  + \frac{i\sigma^{\mu k}}{M} \, A^F_{5}
  + \frac{i\sigma^{\mu \Delta}}{M} \, A^F_{6}
  + \frac{i\sigma^{\mu N}}{M} \, A^F_{7}
  \nonumber\\
 &\quad\hspace{2.55ex}
  + \frac{P^\mu \, i\sigma^{k\Delta}}{M^3} \, A^F_{8}
  + \frac{k^\mu \, i\sigma^{k\Delta}}{M^3} \, A^F_{9}
  + \frac{N^\mu \, i\sigma^{k\Delta}}{M^3} \, A^F_{10}
  + \frac{P^\mu \, i\sigma^{kN}}{M^3} \, A^F_{11}
  + \frac{k^\mu \, i\sigma^{kN}}{M^3} \, A^F_{12}
  \nonumber\\
 &\quad\hspace{2.55ex}
  + \frac{N^\mu \, i\sigma^{kN}}{M^3} \, A^F_{13}
  + \frac{P^\mu \, i\sigma^{\Delta N}}{M^3} \, A^F_{14}
  + \frac{\Delta^\mu \, i\sigma^{\Delta N}}{M^3} \, A^F_{15}
  + \frac{N^\mu \, i\sigma^{\Delta N}}{M^3} \, A^F_{16}
 \bigg] \, u(p, \lambda) \label{e:pvdb} \,,
\end{align}
where the GPCFs $A^F_n$ are scalar functions of $P$, $k$, $\Delta$, $N$, and $\eta$.

%
%
%
\subsection{Parameterization of the axial vector Dirac bilinear}
A complete parameterization of the axial vector Dirac bilinear in eq.~(\ref{e:adb}) can be obtained by treating all possible Dirac currents one after the other:
\begin{enumerate}
\item {\it vector current} [$\bar{u}(p', \lambda') \, \gamma^\mu \, u(p, \lambda)$]: Using the Gordon identity in eq.~(\ref{e:gi1}) all vector currents can be replaced by scalar and tensor currents.
\item {\it axial vector current} [$\bar{u}(p', \lambda') \, \gamma^\mu \gamma_5 \, u(p, \lambda)$]: Using the Gordon identity in eq.~(\ref{e:gi3}) all axial vector currents can be replaced by pseudoscalar and pseudotensor currents.
\item {\it pseudoscalar current} [$\bar{u}(p', \lambda') \, \gamma_5 \, u(p, \lambda)$]: Using eq.~(\ref{e:gi4a}) all pseudoscalar currents can be replaced by pseudotensor currents.
\item {\it tensor current} [$\bar{u}(p', \lambda') \, i\sigma^{\mu\nu} \, u(p, \lambda)$]: Using the $\sigma$-identity in eq.~(\ref{e:si}) all tensor currents can be replaced by pseudotensor currents.
\item {\it pseudotensor current} [$\bar{u}(p', \lambda') \, i\sigma^{\mu\nu} \gamma_5 \, u(p, \lambda)$]: All possible pseudotensor currents are of the form
\begin{equation}
 \bar{u}(p', \lambda') \, i\sigma^{\mu a} \gamma_5 \, u(p, \lambda) \,,\quad
 \bar{u}(p', \lambda') \, a^\mu \, i\sigma^{bc} \gamma_5 \, u(p, \lambda)
\end{equation}
with $a$, $b$ and $c$ being any of the vectors $P$, $k$, $\Delta$, and $N$. Now, contracting eq.~(\ref{e:gi2a}) with $N_\nu$ and in turn with $P_\rho$, $k_\rho$, and $\Delta_\rho$ yields
\begin{align}
 &\bar{u}(p', \lambda') \, i\sigma^{\mu P} \gamma_5 \, u(p, \lambda) \nonumber\\*
 &\quad = \bar{u}(p', \lambda') \, \bigg[
   \frac{P^2 \, i\sigma^{\mu N} \gamma_5}{M^2}
   - \frac{P^\mu \, i\sigma^{PN} \gamma_5}{M^2}
   - \frac{i\varepsilon^{\mu P\Delta N}}{2M^2}
  \bigg] \, u(p, \lambda) \label{e:gi2e} \,,\\
 &\bar{u}(p', \lambda') \, i\sigma^{\mu k} \gamma_5 \, u(p, \lambda) \nonumber\\*
 &\quad = \bar{u}(p', \lambda') \, \bigg[
   \frac{P \cdot k \, i\sigma^{\mu N} \gamma_5}{M^2}
   - \frac{P^\mu \, i\sigma^{kN} \gamma_5}{M^2}
   - \frac{i\varepsilon^{\mu k\Delta N}}{2M^2}
  \bigg] \, u(p, \lambda) \label{e:gi2f} \,,\\
 &\bar{u}(p', \lambda') \, i\sigma^{\mu\Delta} \gamma_5 \, u(p, \lambda) \nonumber\\*
 &\quad = \bar{u}(p', \lambda') \, \bigg[
   - \frac{P^\mu \, i\sigma^{\Delta N} \gamma_5}{M^2}
  \bigg] \, u(p, \lambda) \label{e:gi2g} \,,
\end{align}
which together with eqs.~(\ref{e:gi2b})--(\ref{e:gi2d}) allows to replace some pseudotensor currents by scalar and pseudotensor currents. Therefore, one is left with 13 pseudotensor currents
\begin{align}
 &\bar{u}(p', \lambda') \, i\sigma^{\mu P} \gamma_5 \, u(p, \lambda) \,, &
 &\bar{u}(p', \lambda') \, i\sigma^{\mu k} \gamma_5 \, u(p, \lambda) \,, &
 &\bar{u}(p', \lambda') \, i\sigma^{\mu N} \gamma_5 \, u(p, \lambda) \,, \nonumber\\*
 &\bar{u}(p', \lambda') \, a^\mu \, i\sigma^{PN} \gamma_5 \, u(p, \lambda) \,, &
 &\bar{u}(p', \lambda') \, a^\mu \, i\sigma^{kN} \gamma_5 \, u(p, \lambda) \,, \nonumber\\*
 &\bar{u}(p', \lambda') \, P^\mu \, i\sigma^{\Delta N} \gamma_5 \, u(p, \lambda) \,, &
 &\bar{u}(p', \lambda') \, a^\mu \, i\sigma^{\Delta N} \gamma_5 \, u(p, \lambda) \label{e:adb1}
\end{align}
with $a$ being any of the vectors $k$, $\Delta$, and $N$.
\item {\it scalar current} [$\bar{u}(p', \lambda') \, u(p, \lambda)$]: There are four possible scalar currents
\begin{equation}
 \bar{u}(p', \lambda') \, i\varepsilon^{\mu abc} \, u(p, \lambda) \label{e:adb2}
\end{equation}
with $a$, $b$, and $c$ being any of the vectors $P$, $k$, $\Delta$, and $N$, which can not be replaced.
\end{enumerate}
So far, we were able to reduce the number of currents needed to parameterize the axial vector Dirac bilinear in eq.~(\ref{e:adb}) to the 17 currents in eqs.~(\ref{e:adb1}) and~(\ref{e:adb2}). However, it is possible to reduce this number even further by considering eq.~(\ref{e:di}). Contracting this determinant with $P_\alpha$, $k_\beta$, $\Delta_\gamma$, $N_\delta$, $P_\nu$, $k_\rho$, $\Delta_\sigma$, $N_\tau$ and $i{\sigma_\varepsilon}^N \gamma_5$ and using eqs.~(\ref{e:gi2b})--(\ref{e:gi2d}) and~(\ref{e:gi2e})--(\ref{e:gi2g}) yields
\begin{align}
 &\bar{u}(p', \lambda') \, \det\left(\begin{array}{ccccc}
   0         & \ k^\mu \          & \Delta^\mu     & \ N^\mu \          & 0 \\
   P^2       & \ P \cdot k \      & 0              & \ P \cdot N \      & i\sigma^{PN} \gamma_5 \\
   P \cdot k & \ k^2 \            & k \cdot \Delta & \ k \cdot N \      & i\sigma^{kN} \gamma_5 \\
   0         & \ k \cdot \Delta \ & \Delta^2       & \ \Delta \cdot N \ & i\sigma^{\Delta N} \gamma_5 \\
   P \cdot N & \ k \cdot N \      & \Delta \cdot N & \ 0 \              & 0
  \end{array}\right) \, u(p, \lambda) \nonumber\\
 &\quad = \bar{u}(p', \lambda') \, \det\left(\begin{array}{cccc}
   P \cdot k      & \ 0 \              & P \cdot N      &
     \ - P \cdot N \, i\sigma^{\mu P} \gamma_5 - \tfrac{1}{2} i\varepsilon^{\mu P\Delta N} \\
   k^2            & \ k \cdot \Delta \ & k \cdot N      &
     \ - P \cdot N \, i\sigma^{\mu k} \gamma_5 - \tfrac{1}{2} i\varepsilon^{\mu k\Delta N} \\
   k \cdot \Delta & \ \Delta^2 \       & \Delta \cdot N & \ P^\mu \, i\sigma^{\Delta N} \gamma_5 \\
   k \cdot N      & \ \Delta \cdot N \ & 0              & \ - P \cdot N \, i\sigma^{\mu N} \gamma_5
  \end{array}\right) \, u(p, \lambda) \label{e:adb3} \,.
\end{align}
For $P\not=0$, $k\not=0$, $\Delta\not=0$, and $N\not=0$ this equation is non-trivial and allows one to eliminate one of the 17 currents we have left. As $P$ and $N$ are always different from zero, it is most convenient to eliminate the tensor current
\begin{equation}
 k^\mu \, i\sigma^{\Delta N} \gamma_5
\end{equation}
because either $k$ and $\Delta$ are different form zero and we are able to eliminate it using the constraint in eq.~(\ref{e:adb3}) or at least one of the vectors is zero and the whole current vanishes. Therefore, one possible parameterization of the axial vector Dirac bilinear in eq.~(\ref{e:adb}), is given by
\begin{align}
 &\bar{u}(p', \lambda') \, \Gamma^\mu_\text{A}(P, k, \Delta, N; \eta) \, u(p, \lambda) \nonumber\\
 &\quad = \bar{u}(p', \lambda') \, \bigg[
  \frac{i\varepsilon^{\mu Pk\Delta}}{M^3} \, A^G_{1}
  + \frac{i\varepsilon^{\mu PkN}}{M^3} \, A^G_{2}
  + \frac{i\varepsilon^{\mu P\Delta N}}{M^3} \, A^G_{3}
  + \frac{i\varepsilon^{\mu k\Delta N}}{M^3} \, A^G_{4}
  \nonumber\\
 &\quad\hspace{2.55ex}
  + \frac{i\sigma^{\mu P} \gamma_5}{M} \, A^G_{5}
  + \frac{i\sigma^{\mu k} \gamma_5}{M} \, A^G_{6}
  + \frac{i\sigma^{\mu N} \gamma_5}{M} \, A^G_{7}
  + \frac{k^\mu \, i\sigma^{PN} \gamma_5}{M^3} \, A^G_{8}
  + \frac{\Delta^\mu \, i\sigma^{PN} \gamma_5}{M^3} \, A^G_{9}
  \nonumber\\
 &\quad\hspace{2.55ex}
  + \frac{N^\mu \, i\sigma^{PN} \gamma_5}{M^3} \, A^G_{10}
  + \frac{k^\mu \, i\sigma^{kN} \gamma_5}{M^3} \, A^G_{11}
  + \frac{\Delta^\mu \, i\sigma^{kN} \gamma_5}{M^3} \, A^G_{12}
  + \frac{N^\mu \, i\sigma^{kN} \gamma_5}{M^3} \, A^G_{13}
  \nonumber\\
 &\quad\hspace{2.55ex}
  + \frac{P^\mu \, i\sigma^{\Delta N} \gamma_5}{M^3} \, A^G_{14}
  + \frac{\Delta^\mu \, i\sigma^{\Delta N} \gamma_5}{M^3} \, A^G_{15}
  + \frac{N^\mu \, i\sigma^{\Delta N} \gamma_5}{M^3} \, A^G_{16}
 \bigg] \, u(p, \lambda) \label{e:padb} \,,
\end{align}
where the GPCFs $A^G_n$ are scalar functions of $P$, $k$, $\Delta$, $N$, and $\eta$.

%
%
%
\subsection{Parameterization of the tensor Dirac bilinear}
A complete parameterization of the tensor Dirac bilinear in eq.~(\ref{e:tdb}) can be constructed from the parameterization of the pseudoscalar and the vector Dirac bilinears in eqs.~(\ref{e:ppdb}) and~(\ref{e:pvdb}). Respecting the antisymmetry of the tensor Dirac bilinear in eq.~(\ref{e:tdb}) it can be rewritten as
\begin{align}
 &\bar{u}(p', \lambda') \, \Gamma^{\mu\nu}_\text{T}(P, k, \Delta, N; \eta) \, u(p, \lambda) \nonumber\\
 &\quad = (\delta^\mu_\rho \delta^\nu_\sigma - \delta^\nu_\rho \delta^\mu_\sigma) \, \bar{u}(p', \lambda') \, \bigg[
  \bigg( \frac{P^\rho}{M} - \frac{P^2 \, N^\rho}{2M^3} \bigg) \,
  \frac{M \, \Gamma^{+\sigma}_\text{T}(P, k, \Delta, N; \eta)}{P \cdot n} \nonumber\\*
 &\quad\hspace{2.55ex}
  + \frac{N^\rho}{M} \, \bigg( \delta^\sigma_\tau - \frac{P^\sigma N_\tau}{M^2} \bigg) \,
  \frac{(P \cdot n) \, \Gamma^{-\tau}_\text{T}(P, k, \Delta, N; \eta)}{M} \nonumber\\*
 &\quad\hspace{2.55ex}
  - \frac{i\varepsilon^{\rho\sigma PN}}{2 M^2} \,
  i\Gamma^{12}_\text{T}(P, k, \Delta, N; \eta)
  \bigg] \, u(p, \lambda) \label{e:tdb1} \,,
\end{align}
where from eqs.~(\ref{e:ssp})--(\ref{e:tsp}) it follows that the structures $M \, \Gamma^{+\mu}_\text{T} / (P \cdot n)$ and $(P \cdot n) \, \Gamma^{-\mu}_\text{T} / M$ behave like vectors under parity whereas the structure $i\Gamma^{12}_\text{T}$ behaves like a pseudoscalar. Therefore, inserting the respective parameterization from eqs.~(\ref{e:ppdb}) and~(\ref{e:pvdb}) into~(\ref{e:tdb1}) a possible parameterization of the tensor Dirac bilinear in eq.~(\ref{e:tdb}) is given by
\begin{align}
 &\bar{u}(p', \lambda') \, \Gamma^{\mu\nu}_\text{T}(P, k, \Delta, N; \eta) \, u(p, \lambda) \nonumber\\
 &\quad = (\delta^\mu_\rho \delta^\nu_\sigma - \delta^\nu_\rho \delta^\mu_\sigma) \, \bar{u}(p', \lambda') \, \bigg[
  \bigg( \frac{P^\rho}{M} - \frac{P^2 \, N^\rho}{2M^3} \bigg) \, \bigg(
  \frac{P^\sigma}{M} \, A_{1}
  + \frac{k^\sigma}{M} \, A_{2}
  + \frac{\Delta^\sigma}{M} \, A_{3}
  + \frac{N^\sigma}{M} \, A_{4}
  \nonumber\\
 &\quad\hspace{2.55ex}
  + \frac{i\sigma^{\sigma k}}{M} \, A_{5}
  + \frac{i\sigma^{\sigma \Delta}}{M} \, A_{6}
  + \frac{i\sigma^{\sigma N}}{M} \, A_{7}
  + \frac{P^\sigma \, i\sigma^{k\Delta}}{M^3} \, A_{8}
  + \frac{k^\sigma \, i\sigma^{k\Delta}}{M^3} \, A_{9}
  + \frac{N^\sigma \, i\sigma^{k\Delta}}{M^3} \, A_{10}
  \nonumber\\
 &\quad\hspace{2.55ex}
  + \frac{P^\sigma \, i\sigma^{kN}}{M^3} \, A_{11}
  + \frac{k^\sigma \, i\sigma^{kN}}{M^3} \, A_{12}
  + \frac{N^\sigma \, i\sigma^{kN}}{M^3} \, A_{13}
  + \frac{P^\sigma \, i\sigma^{\Delta N}}{M^3} \, A_{14}
  + \frac{\Delta^\sigma \, i\sigma^{\Delta N}}{M^3} \, A_{15}
  \nonumber\\
 &\quad\hspace{2.55ex}
  + \frac{N^\sigma \, i\sigma^{\Delta N}}{M^3} \, A_{16} \bigg)
  + \frac{N^\rho}{M} \, \bigg( \delta^\sigma_\tau - \frac{P^\sigma N_\tau}{M^2} \bigg) \, \bigg(
  \frac{P^\tau}{M} \, A_{17}
  + \frac{k^\tau}{M} \, A_{18}
  + \frac{\Delta^\tau}{M} \, A_{19}
  + \frac{N^\tau}{M} \, A_{20}
  \nonumber\\
 &\quad\hspace{2.55ex}
  + \frac{i\sigma^{\tau k}}{M} \, A_{21}
  + \frac{i\sigma^{\tau \Delta}}{M} \, A_{22}
  + \frac{i\sigma^{\tau N}}{M} \, A_{23}
  + \frac{P^\tau \, i\sigma^{k\Delta}}{M^3} \, A_{24}
  + \frac{k^\tau \, i\sigma^{k\Delta}}{M^3} \, A_{25}
  + \frac{N^\tau \, i\sigma^{k\Delta}}{M^3} \, A_{26}
  \nonumber\\
 &\quad\hspace{2.55ex}
  + \frac{P^\tau \, i\sigma^{kN}}{M^3} \, A_{27}
  + \frac{k^\tau \, i\sigma^{kN}}{M^3} \, A_{28}
  + \frac{N^\tau \, i\sigma^{kN}}{M^3} \, A_{29}
  + \frac{P^\tau \, i\sigma^{\Delta N}}{M^3} \, A_{30}
  + \frac{\Delta^\tau \, i\sigma^{\Delta N}}{M^3} \, A_{31}
  \nonumber\\
 &\quad\hspace{2.55ex}
  + \frac{N^\tau \, i\sigma^{\Delta N}}{M^3} \, A_{32} \bigg)
  - \frac{i\varepsilon^{\rho\sigma PN}}{2M^2} \, \bigg(
  \frac{i\varepsilon^{Pk\Delta N}}{M^4} \, A_{33}
  + \frac{i\sigma^{PN} \gamma_5}{M^2} \, A_{34}
  + \frac{i\sigma^{kN} \gamma_5}{M^2} \, A_{35}
  \nonumber\\
 &\quad\hspace{2.55ex}
  + \frac{i\sigma^{\Delta N} \gamma_5}{M^2} \, A_{36} \bigg)
 \bigg] \, u(p, \lambda) \label{e:tdb2} \,.
\end{align}
Of course, not all 36 structures in this parameterization are independent.
It is therefore necessary to rearrange them in order to obtain a minimal set of structures for the parameterization of the tensor Dirac bilinear in eq.~(\ref{e:tdb}).
By first performing all possible contractions in eq.~(\ref{e:tdb2}) and after that using the $\sigma$-identity in~(\ref{e:si}) for the last three terms as well as the identity
\begin{equation}
 \epsilon^{\alpha\beta\gamma\delta}\epsilon^{\mu\nu\rho\sigma}
 =-\det\left(\begin{array}{cccc}
  g^{\alpha\mu} & \ g^{\beta\mu} \ & g^{\gamma\mu} & \ g^{\delta\mu} \\
  g^{\alpha\nu} & \ g^{\beta\nu} \ & g^{\gamma\nu} & \ g^{\delta\nu} \\
  g^{\alpha\rho} & \ g^{\beta\rho} \ & g^{\gamma\rho} & \ g^{\delta\rho} \\
  g^{\alpha\sigma} & \ g^{\beta\sigma} \ & g^{\gamma\sigma} & \ g^{\delta\sigma} \\
 \end{array}\right)
\end{equation}
for all products of two $\epsilon$-tensors the number of independent structures already reduces significantly.
Finally, the Gordon identity in eq.~(\ref{e:gi2}) allows one to express all tensor currents containing a $\sigma$-matrix contracted with $P$ in terms of scalar currents.
This reduces the number of independent structures to 24, which are given for example by
\begin{align}
 &\bar{u}(p', \lambda') \, \Gamma^{\mu\nu}_\text{T}(P, k, \Delta, N; \eta) \, u(p, \lambda) \nonumber\\
 &\quad = (\delta^\mu_\rho \delta^\nu_\sigma - \delta^\nu_\rho \delta^\mu_\sigma) \, \bar{u}(p', \lambda') \, \bigg[
  \frac{P^\rho k^\sigma}{M^2} \, A^H_{1}
  + \frac{P^\rho \Delta^\sigma}{M^2} \, A^H_{2}
  + \frac{P^\rho N^\sigma}{M^2} \, A^H_{3}
  + \frac{k^\rho \Delta^\sigma}{M^2} \, A^H_{4}
  \nonumber\\
 &\quad\hspace{2.55ex}
  + \frac{k^\rho N^\sigma}{M^2} \, A^H_{5}
  + \frac{\Delta^\rho N^\sigma}{M^2} \, A^H_{6}
  + i\sigma^{\rho\sigma} \, A^H_{7}
  + \frac{P^\rho \, i\sigma^{\sigma k}}{M^2} \, A^H_{8}
  + \frac{N^\rho \, i\sigma^{\sigma k}}{M^2} \, A^H_{9}
  \nonumber\\
 &\quad\hspace{2.55ex}
  + \frac{P^\rho \, i\sigma^{\sigma\Delta}}{M^2} \, A^H_{10}
  + \frac{N^\rho \, i\sigma^{\sigma\Delta}}{M^2} \, A^H_{11}
  + \frac{P^\rho \, i\sigma^{\sigma N}}{M^2} \, A^H_{12}
  + \frac{k^\rho \, i\sigma^{\sigma N}}{M^2} \, A^H_{13}
  + \frac{\Delta^\rho \, i\sigma^{\sigma N}}{M^2} \, A^H_{14}
  \nonumber\\
 &\quad\hspace{2.55ex}
  + \frac{N^\rho \, i\sigma^{\sigma N}}{M^2} \, A^H_{15}
  + \frac{P^\rho k^\sigma \, i\sigma^{k\Delta}}{M^4} \, A^H_{16}
  + \frac{P^\rho N^\sigma \, i\sigma^{k\Delta}}{M^4} \, A^H_{17}
  + \frac{k^\rho N^\sigma \, i\sigma^{k\Delta}}{M^4} \, A^H_{18}
  \nonumber\\
 &\quad\hspace{2.55ex}
  + \frac{P^\rho k^\sigma \, i\sigma^{kN}}{M^4} \, A^H_{19}
  + \frac{P^\rho N^\sigma \, i\sigma^{kN}}{M^4} \, A^H_{20}
  + \frac{k^\rho N^\sigma \, i\sigma^{kN}}{M^4} \, A^H_{21}
  + \frac{P^\rho \Delta^\sigma \, i\sigma^{\Delta N}}{M^4} \, A^H_{22}
  \nonumber\\
 &\quad\hspace{2.55ex}
  + \frac{P^\rho N^\sigma \, i\sigma^{\Delta N}}{M^4} \, A^H_{23}
  + \frac{\Delta^\rho N^\sigma \, i\sigma^{\Delta N}}{M^4} \, A^H_{24}
 \bigg] \, u(p, \lambda) \label{e:ptdb} \,,
\end{align}
where the GPCFs $A^H_n$ are scalar functions of $P$, $k$, $\Delta$, $N$, and $\eta$ and are related to the structures $A_n$ in eq.~(\ref{e:tdb2}) by
\begin{eqnarray}
 A^H_{1}
 &=& A_{2} + \frac{\Delta \cdot N}{M^2} \, A_{33} \,, \\
 A^H_{2}
 &=& A_{3} - \frac{k \cdot N}{M^2} \, A_{33} \,, \\
 A^H_{3}
 &=& \frac{P^2}{2M^2} \, A_{1} + A_{4} + \frac{k \cdot N}{M^2} \, A_{18}
     + \frac{\Delta \cdot N}{M^2} \, A_{19} \nonumber\\*
 & & - \frac{(P \cdot k) \, (\Delta \cdot N)}{M^4} \, A_{33}
     - \frac{\Delta \cdot N}{2 M^2} \, A_{34} \,, \\
 A^H_{4}
 &=& A_{33} \,, \\
 A^H_{5}
 &=& \frac{P^2}{2M^2} \, A_{2} - A_{18}
     + \frac{P^2 \, (\Delta \cdot N)}{M^4} \, A_{33}
     - \frac{\Delta \cdot N}{2 M^2} \, A_{35} \,, \\
 A^H_{6}
 &=& \frac{P^2}{2M^2} \, A_{3} - A_{19}
     - \frac{P^2 \, (k \cdot N) - (P \cdot k) \, (P \cdot N)}{M^4} \, A_{33} \nonumber\\*
 & & + \tfrac{1}{2} A_{34} + \frac{k \cdot N}{2 M^2} \, A_{35} \,, \\
 A^H_{7}
 &=& \tfrac{1}{2} A_{34} + \frac{k \cdot N}{2 M^2} \, A_{35}
     + \frac{\Delta \cdot N}{2 M^2} \, A_{36} \,, \\
 A^H_{8}
 &=& A_{5} \,, \\
 A^H_{9}
 &=& - \frac{P^2}{2M^2} \, A_{5} + A_{21} \,, \\
 A^H_{10}
 &=& A_{6} \,, \\
 A^H_{11}
 &=& - \frac{P^2}{2M^2} \, A_{6} + A_{22} \,, \\
 A^H_{12}
 &=& A_{7} + A_{34} \,, \\
 A^H_{13}
 &=& A_{35} \,, \\
 A^H_{14}
 &=& A_{36} \,, \\
 A^H_{15}
 &=& - \frac{P^2}{2M^2} \, A_{7} + A_{23} - \frac{P^2}{M^2} \, A_{34}
     - \frac{P \cdot k}{M^2} \, A_{35} \,, \\
 A^H_{16}
 &=& A_{9} \,, \\
 A^H_{17}
 &=& \frac{P^2}{2M^2} \, A_{8} + A_{10} + \frac{k \cdot N}{M^2} \, A_{25} \,, \\
 A^H_{18}
 &=& \frac{P^2}{2M^2} \, A_{9} - A_{25} \,, \\
 A^H_{19}
 &=& A_{12} \,, \\
 A^H_{20}
 &=& \frac{P^2}{2M^2} \, A_{11} + A_{13} - A_{21}
     + \frac{k \cdot N}{M^2} \, A_{28} \,, \\
 A^H_{21}
 &=& \frac{P^2}{2M^2} \, A_{12} - A_{28} \,, \\
 A^H_{22}
 &=& A_{15} \,, \\
 A^H_{23}
 &=& \frac{P^2}{2M^2} \, A_{14} + A_{16} - A_{22}
     + \frac{\Delta \cdot N}{M^2} \, A_{31} \,, \\
 A^H_{24}
 &=& \frac{P^2}{2M^2} \, A_{15} - A_{31} \,.
\end{eqnarray}

%
%
%
\section{Relations between GTMDs and GPCFs}\label{c:app_gtmd_gpcf}
Here the explicit relations between the leading twist GTMDs in 
eqs.~(\ref{e:gtmd_1})--(\ref{e:gtmd_3}) and the GPCFs in 
eqs.~(\ref{e:wv_res})--(\ref{e:wt_res}) are listed. 
For brevity we leave out the arguments of the functions. 
Straightforward calculation leads to the results
\begin{eqnarray}
 F_{1,1}
 &=& 2P^+ \, \int dk^- \, \bigg[
     A^F_{1} + x A^F_{2} - 2\xi A^F_{3}
     + \xi \, \bigg(x A^F_{5} - 2\xi A^F_{6}\bigg) \nonumber\\*
 & & \hspace{14.15ex}
     + \frac{2\xi \, k \cdot \Delta + x \Delta^2}{2 M^2} \, \bigg(A^F_{8} + x A^F_{9}\bigg) \nonumber\\*
 & & \hspace{14.15ex}
     - x\xi \, \bigg(A^F_{11} + x A^F_{12}\bigg)
     + 2\xi^2 \, \bigg(A^F_{14} - 2\xi A^F_{15}\bigg)
     \bigg] \,, \label{e:gtmd_gpcf_1} \\
 F_{1,2}
 &=& 2P^+ \, \int dk^- \, \bigg[
     A^F_{5}
     - \frac{2\xi P^2}{M^2} \, \bigg(A^F_{8} + x A^F_{9}\bigg)
     - \bigg(A^F_{11} + x A^F_{12}\bigg)
     \bigg] \,, \label{e:gtmd_gpcf_2} \\
 F_{1,3}
 &=& 2P^+ \, \int dk^- \, \bigg[
     A^F_{6}
     - \frac{x P^2 - P \cdot k}{M^2} \, \bigg(A^F_{8} + x A^F_{9}\bigg)
     - \bigg(A^F_{14} - 2\xi A^F_{15}\bigg)
     \bigg] \,, \label{e:gtmd_gpcf_3} \\
 F_{1,4}
 &=& 2P^+ \, \int dk^- \, \bigg[
     A^F_{8} + x A^F_{9}
     \bigg] \,, \label{e:gtmd_gpcf_4} \\
 G_{1,1}
 &=& 2P^+ \, \int dk^- \, \bigg[
     A^G_{1}
     \bigg] \,, \label{e:gtmd_gpcf_5} \\
 G_{1,2}
 &=& 2P^+ \, \int dk^- \, \bigg[
     A^G_{6}
     - \bigg(x A^G_{11} - 2\xi A^G_{12}\bigg)
     \bigg] \,, \label{e:gtmd_gpcf_6} \\
 G_{1,3}
 &=& 2P^+ \, \int dk^- \, \bigg[
     - \bigg(A^G_{14} - 2\xi A^G_{15}\bigg)
     \bigg] \,, \label{e:gtmd_gpcf_7} \\
 G_{1,4}
 &=& 2P^+ \, \int dk^- \, \bigg[
     A^G_{5} + x A^G_{6} - x A^G_{8} + 2\xi A^G_{9}
     - x \, \bigg(x A^G_{11} - 2\xi A^G_{12}\bigg) \nonumber\\*
 & & \hspace{14.15ex}
     + 2\xi \, \bigg(A^G_{14} - 2\xi A^G_{15}\bigg)
     \bigg] \,, \label{e:gtmd_gpcf_8} \\
 H_{1,1}
 &=& 2P^+ \, \int dk^- \, \bigg[
     A^H_{1} + 2\xi A^H_{4}
     + \frac{2\xi \, k \cdot \Delta + x \Delta^2}{2M^2} \, A^H_{16}
     - x\xi A^H_{19}
     \bigg] \,, \label{e:gtmd_gpcf_9} \\
 H_{1,2}
 &=& 2P^+ \, \int dk^- \, \bigg[
     A^H_{2} + x A^H_{4} - \tfrac{1}{2}x A^H_{8} + \xi A^H_{10} + 2\xi^2 A^H_{22}
     \bigg] \,, \label{e:gtmd_gpcf_10} \\
 H_{1,3}
 &=& 2P^+ \, \int dk^- \, \bigg[
     2 A^H_{7}
     + \frac{x P^2 - P \cdot k}{M^2} \, A^H_{8}
     - \frac{2\xi P^2}{M^2} \, A^H_{10}
     - \bigg(A^H_{12} + x A^H_{13} - 2\xi A^H_{14}\bigg) \nonumber\\*
 & & \hspace{14.15ex}
     - \frac{P^2 \, (2\xi k^2 + x \, k \cdot \Delta)
       - P \cdot k \, (2\xi \, P \cdot k + k \cdot \Delta)}{M^4} \, A^H_{16} \nonumber\\*
 & & \hspace{14.15ex}
     - \frac{x^2 P^2 - 2x \, P \cdot k + k^2}{M^2} \, A^H_{19}
     - \frac{4\xi^2 P^2 + \Delta^2}{M^2} \, A^H_{22}
     \bigg] \,, \label{e:gtmd_gpcf_11} \\
 H_{1,4}
 &=& 2P^+ \, \int dk^- \, \bigg[
     - A^H_{19}
     - \frac{2\xi P^2}{M^2} \, A^H_{16}
     \bigg] \,, \label{e:gtmd_gpcf_12} \\
 H_{1,5}
 &=& 2P^+ \, \int dk^- \, \bigg[
     - \frac{x P^2 - P \cdot k}{M^2} \, A^H_{16}
     \bigg] \,, \label{e:gtmd_gpcf_13} \\
 H_{1,6}
 &=& 2P^+ \, \int dk^- \, \bigg[
     - A^H_{22}
     \bigg] \,, \label{e:gtmd_gpcf_14} \\
 H_{1,7}
 &=& 2P^+ \, \int dk^- \, \bigg[
     - A^H_{8}
     - \frac{2x\xi P^2 - 2\xi \, P \cdot k - k \cdot \Delta}{M^2} \, A^H_{16}
     \bigg] \,, \label{e:gtmd_gpcf_15} \\
 H_{1,8}
 &=& 2P^+ \, \int dk^- \, \bigg[
     - A^H_{10}
     - \frac{x^2 P^2 - 2x \, P \cdot k + k^2}{M^2} \, A^H_{16}
     \bigg] \,. \label{e:gtmd_gpcf_16}
\end{eqnarray}

%
%
%
\section{Model calculation of GTMDs}\label{c:app_gtmd_model}
For illustrative purposes and in order to get a first estimate we calculate all 
the leading twist GTMDs in the scalar spectator diquark model of the nucleon
(see, e.g., ref.~\cite{Brodsky:2002cx}) by restricting ourselves to lowest nontrivial 
order in perturbation theory. 
The Lagrangian of the diquark model reads
\begin{eqnarray}
\mathcal{L_\text{SDM}}(x) & = &
 \bar\Psi(x)\,(i\gamma^\mu\partial_\mu-M)\,\Psi(x)
+\bar\psi(x)\,(i\gamma^\mu\partial_\mu-m_q)\,\psi(x)
\nonumber \\
& & +\partial^{\mu} \varphi^*(x) \, \partial_{\mu} \varphi(x) 
 -m_s^2 \, \varphi^*(x) \varphi(x)
\nonumber \\
& & +g\,\big[\bar\psi(x)\,\Psi(x)\,\varphi^*(x)+\bar\Psi(x)\,\psi(x)\,\varphi(x)\big] \,,
\end{eqnarray}
where $\Psi$ denotes the nucleon field, $\psi$ the quark field, and $\varphi$ 
the scalar diquark field. 
The essential ingredient of the model is a three-point interaction between the target, 
quarks, and diquarks, with the coupling constant $g$.
This framework allows one to carry out perturbative calculations.
All the results for parton distributions given below contain the coupling $g$ to
the second power.
Notice also that the condition $M < m_q + m_s$ has to be fulfilled in order to have 
a stable target state.
%
\FIGURE[t]{%
 \includegraphics{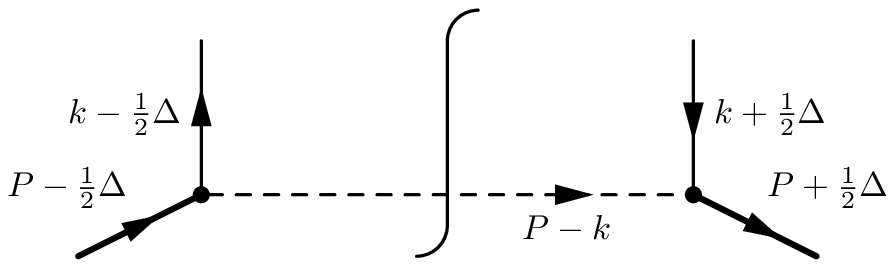}
 \includegraphics{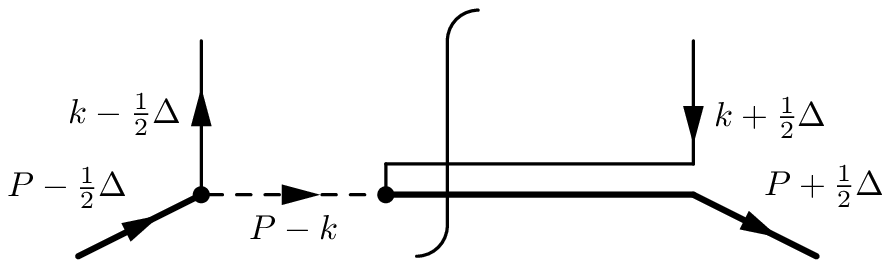}
 \includegraphics{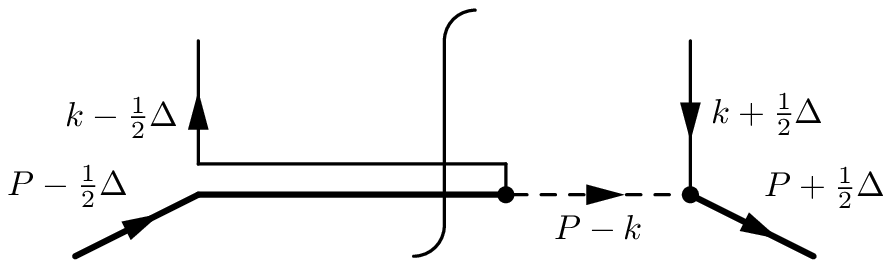}
 \caption{Lowest nontrivial order diagrams contributing to the GTMDs in the scalar 
spectator diquark model.
The second and third diagram are only relevant for the ERBL region which is 
characterized by $|x| \le |\xi|$.}
 \label{f:model}}
%

The lowest order contribution to the generalized $k_T$-dependent correlator in 
eq.~(\ref{e:corr_gtmd}) comes from the tree-level diagrams depicted in 
figure~\ref{f:model}. 
These diagrams can be evaluated in a straightforward manner yielding 
\begin{align} \label{e:corr_model}
 & W_{\lambda\lambda'}^{[\Gamma]}(P,x,\vec{k}_T,\Delta,N;\eta) \nonumber \\
 & \quad = \frac{g^2 \, (1-x) \, (x^2-\xi^2)}{4(2\pi)^3 P^+} \, \bigg[
    \frac{N(k_1^-)}{D_+ \, D_-} \, \Theta(1-x)
    - \frac{N(k_2^-)}{D_+ \, (D_- - D_+)} \, \Theta(\xi-x) \nonumber\\
 & \quad\qquad - \frac{N(k_3^-)}{(D_+ - D_-) \, D_-} \, \Theta(-\xi-x)
   \bigg] \Theta(1-|\xi|) \,,
\end{align}
where the numerator $N(k^-)$ is given by
\begin{equation} \label{e:num_model}
 N(k^-) = \bar{u}(p', \lambda') \, (\kslash + \tfrac{1}{2} \Dslash + m_q) \, \Gamma \,
  (\kslash - \tfrac{1}{2} \Dslash + m_q) \, u(p, \lambda) \,,
\end{equation}
the denominators $D_\pm$ by
\begin{align}
 D_\pm
 &= \frac{x\pm\xi}{1\mp\xi} \, \bigg( (1\mp\xi) \, \vec{k}_T \pm \tfrac{1}{2}(1-x) \, \vec{\Delta}_T \bigg)^2 
    + (x^2-\xi^2) \, m_s^2 \nonumber\\
 &\quad + (1-x) \, (x\pm\xi) \, m_q^2 - \frac{(1-x) \, (x^2-\xi^2)}{1\mp\xi} \, M^2 \,,
\end{align}
and $k^-$ is fixed by the cuts in the diagrams to be
\begin{align}
 k_1^-
 &= \frac{\tfrac{1}{4} \vec{\Delta}_T^2 + M^2}{2(1-\xi^2)P^+} 
   - \frac{\vec{k}_T^2 + m_s^2}{2(1-x)P^+} \,,\\
 k_2^-
 &= -\xi \, \frac{\tfrac{1}{4} \vec{\Delta}_T^2 + M^2}{2(1-\xi^2)P^+} 
   + \frac{(\vec{k}_T + \tfrac{1}{2}\vec{\Delta}_T)^2 + m_q^2}{2(x-\xi)P^+} \,,\\
 k_3^-
 &= \xi \, \frac{\tfrac{1}{4} \vec{\Delta}_T^2 + M^2}{2(1-\xi^2)P^+} 
   + \frac{(\vec{k}_T - \tfrac{1}{2}\vec{\Delta}_T)^2 + m_q^2}{2(x+\xi)P^+} \,.
\end{align}
Since the calculation is carried out only to lowest order in perturbation theory, 
no effect due to the Wilson line enters. 
As a consequence, the correlator~(\ref{e:corr_model}) actually does not depend on 
the parameter $\eta$.

Using now the expression~(\ref{e:corr_model}) and the definitions for the GTMDs in 
eqs.~(\ref{e:gtmd_1})--(\ref{e:gtmd_3}) it is possible to calculate the leading 
twist GTMDs. 
As for leading twist the numerator $N(k^-)$ in eq.~(\ref{e:num_model}) does actually 
not depend on $k^-$, the final expressions are not too complicated.
They read\footnote{In ref.~\cite{Meissner:2008ay} we obtained results for the GTMDs 
of a spin-0 target in a simple spectator model.
Those results are only complete for $|\xi| \le x \le 1$.
But note also that this specific kinematical region includes the case $\xi = 0$ which
is relevant for discussing potential nontrivial relations between GPDs and TMDs.}
\begin{eqnarray}
 F_{1,1}^{e}
 &=& C \, \bigg[ (1-\xi^2) \, \bigg(\vec{k}_T^2 + (m_q + M)^2\bigg)
     + (1-x) \, \bigg(\xi \, \vec{k}_T \cdot \vec{\Delta}_T - 2 (m_q + x M) M\bigg) \nonumber\\*
 & & \quad\qquad - \tfrac{1}{4}(1-x)^2 \, \bigg(\vec{\Delta}_T^2 + 4M^2\bigg) \bigg]
 \,, \label{e:gtmd_model_1} \\
 F_{1,2}^{e}
 &=& 2C \, \bigg[ \frac{\xi(1-x)}{4(1-\xi^2)} \, \bigg(\vec{\Delta}_T^2 + 4M^2\bigg) \bigg]
 \,, \label{e:gtmd_model_2} \\
 F_{1,3}^{e}
 &=& \tfrac{1}{2}C \, \bigg[ \bigg(\vec{k}_T^2 + (m_q + M)^2\bigg)
     - \frac{(1-x)^2}{4(1-\xi^2)} \, \bigg(\vec{\Delta}_T^2 + 4M^2\bigg) \bigg]
 \,, \label{e:gtmd_model_3} \\
 F_{1,4}^{e}
 &=& -C \, \bigg[ (1-x) \, M^2 \bigg]
 \,, \label{e:gtmd_model_4} \\
 G_{1,1}^{e}
 &=& -C \, \bigg[ (1-x) \, M^2 \bigg]
 \,, \label{e:gtmd_model_5} \\
 G_{1,2}^{e}
 &=& -2C \, \bigg[ \frac{1-x}{4(1-\xi^2)} \, \bigg(\vec{\Delta}_T^2 + 4M^2\bigg)
     -(m_q + M) M \bigg]
 \,, \label{e:gtmd_model_6} \\
 G_{1,3}^{e}
 &=& -\tfrac{1}{2}C \, \bigg[ \xi \, \bigg(\vec{k}_T^2 - (m_q + M)^2\bigg)
     -(1-x) \, \vec{k}_T \cdot \vec{\Delta}_T \nonumber\\*
 & & - \frac{\xi(1-x)^2}{4(1-\xi^2)} \, \bigg(\vec{\Delta}_T^2 + 4M^2\bigg) \bigg]
 \,, \label{e:gtmd_model_7} \\
 G_{1,4}^{e}
 &=& -C \, \bigg[ (1-\xi^2) \, \bigg(\vec{k}_T^2 - (m_q + M)^2\bigg)
     + (1-x) \, \bigg(\xi \, \vec{k}_T \cdot \vec{\Delta}_T + 2 (m_q + M) M\bigg) \nonumber\\*
 & & \quad\qquad - \tfrac{1}{4}(1-x)^2 \, \bigg(\vec{\Delta}_T^2 + 4M^2\bigg) \bigg]
 \,, \label{e:gtmd_model_8} \\
 H_{1,1}^{e}
 &=& 2C \, \bigg[ \xi (1-x) \, M^2 \bigg]
 \,, \label{e:gtmd_model_9} \\
 H_{1,2}^{e}
 &=& C \, \bigg[ (1-x) \, (m_q + x M) M \bigg]
 \,, \label{e:gtmd_model_10} \\
 H_{1,3}^{e}
 &=& C \, \bigg[ \bigg(\vec{k}_T^2 + (m_q + M)^2\bigg)
     + \frac{\xi \, (m_q + M)}{M} \,
       \frac{\vec{k}_T^2 \vec{\Delta}_T^2 - (\vec{k}_T \cdot \vec{\Delta}_T)^2}{\vec{k}_T \cdot \vec{\Delta}_T}
     \nonumber\\*
 & & \quad\qquad - \frac{(1-x) \, ((m_q + M) + (m_q + x M))}{4(1-\xi^2) \, M} \,
       \bigg(\vec{\Delta}_T^2 + 4M^2\bigg) \bigg]
 \,, \label{e:gtmd_model_11} \\
 H_{1,4}^{e}
 &=& -C \, \bigg[ \frac{\xi \, \vec{\Delta}_T^2}{\vec{k}_T \cdot \vec{\Delta}_T} \, (m_q + M) M + 2 M^2 \bigg]
 \,, \label{e:gtmd_model_12} \\
 H_{1,5}^{e}
 &=& C \, \bigg[ \xi \, (m_q + M) M \bigg]
 \,, \label{e:gtmd_model_13} \\
 H_{1,6}^{e}
 &=& -C \, \bigg[ \bigg( \frac{\xi \, \vec{k}_T^2}{\vec{k}_T \cdot \vec{\Delta}_T} - \tfrac{1}{2}(1-x) \bigg) \,
     (m_q + M) M \bigg]
 \,, \label{e:gtmd_model_14} \\
 H_{1,7}^{e}
 &=& -2C \, \bigg[ (1-\xi^2) \, (m_q + M) M - (1-x) \, M^2 \bigg]
 \,, \label{e:gtmd_model_15} \\
 H_{1,8}^{e}
 &=& -C \, \bigg[ \xi (1-x) \, (m_q + M) M \bigg]
 \,, \label{e:gtmd_model_16}
\end{eqnarray}
with\footnote{Note that in the context of GPDs one typically only considers the case 
$\xi \ge 0$ as this condition is satisfied for all known processes where GPDs can be
measured.
Here we do not use this restriction and take negative values of $\xi$ into account 
as well.}
\begin{equation}
 C = \left\{
  \begin{array}{ccl}
   0 & \phantom{\Bigg[} & \text{for $x\ge1$ (unphysical region),}\\
   \displaystyle\frac{g^2 \, (1-x) \, (x^2-\xi^2)}{2(2\pi)^3 \, D_+ \, D_-} & \phantom{\bigg[}
   & \text{for $|\xi|\le x\le1$ (DGLAP region for quarks),} \\
   \displaystyle\frac{g^2 \, (1-x) \, (x^2-\xi^2)}{2(2\pi)^3 \, (D_+ - D_-) \, D_-} & \phantom{\Bigg[}
   & \text{for $-1\le-\xi\le x\le\xi\le1$ (ERBL region for $\xi\ge0$),} \\
   \displaystyle\frac{g^2 \, (1-x) \, (x^2-\xi^2)}{2(2\pi)^3 \, D_+ \, (D_- - D_+)} & \phantom{\bigg[}
   & \text{for $-1\le\xi\le x\le-\xi\le1$ (ERBL region for $\xi\le0$),} \\
   0 & \phantom{\bigg[} & \text{for $-1\le x\le-|\xi|$ (DGLAP region for antiquarks),} \\
   0 & \phantom{\bigg[} & \text{for $x\le-1$ (unphysical region).}
  \end{array}
 \right.
\end{equation}
To shorten the notation we have suppressed the arguments of the GTMDs. 
All (na\"ive) T-odd parts of GTMDs vanish to lowest order in perturbation theory 
investigated here. 
To get nonzero results for these functions requires considering at least one-loop corrections 
that include effects from the Wilson line. 
We checked that the limits of TMDs and GPDs for $\xi=0$ agree with the results found 
in \cite{Meissner:2007rx}.

%
%
%
\bibliographystyle{JHEP}
\bibliography{gtmd_2}
\end{document}